\documentclass[aps,prd,preprint,superscriptaddress,showpacs,nofootinbib,floatfix]{revtex4-1}
\pdfoutput=1

\usepackage{amsmath}
\usepackage{amssymb}
\usepackage{mathtools}
\usepackage{graphicx}
\usepackage{relsize}
\usepackage[hidelinks]{hyperref}
\usepackage{wasysym}

\usepackage{color}

\begin{document}


\title{Stochastic dark energy from inflationary quantum fluctuations}


\author{Dra\v{z}en~Glavan}
\email[]{drazen.glavan@fuw.edu.pl}
\affiliation{Institute of Theoretical Physics, Faculty of Physics, 
University of Warsaw, Pasteura 5, 02-093 Warsaw, Poland}

\author{Tomislav~Prokopec}
\email[]{t.prokopec@uu.nl}
\affiliation{Institute for Theoretical Physics, Spinoza Institute
$\&$ EMME$\Phi$, Faculty of Science, Utrecht University,
Postbus 80.195, 3508 TD Utrecht, The Netherlands}

\author{Alexei~A.~Starobinsky}
\email[]{alstar@landau.ac.ru}
\affiliation{L.~D.~Landau Institute for Theoretical Physics RAS, 
Moscow 119334, Russian Federation}
\affiliation{Kazan Federal University, Kazan 420008, 
Republic of Tatarstan, Russian Federation
\\
\
\\
\
}


\begin{abstract}

We study the quantum backreaction from inflationary fluctuations of 
a very light, non-minimally coupled spectator scalar and show that it 
is a viable candiate for dark energy. The problem is solved by suitably 
adapting the formalism of stochastic inflation. This allows us to 
self-consistently account for the backreaction on the background
expansion rate of the Universe where its effects are large. This 
framework is equivalent to that of semiclassical gravity in which 
matter vacuum fluctuations are included at the one loop level, but 
purely quantum gravitational fluctuations are neglected.  Our results 
show that dark energy in our model can be characterized by a 
distinct effective equation of state parameter (as a function of redshift) 
which allows for testing of the model at the level of the 
background.

\end{abstract}

\maketitle


\section{Introduction}
\label{sec:Introduction}

The origin of dark energy (DE) is one of the most fascinating 
unsolved problems of modern science. In literature traditionally 
two main classes of solutions have been 
proposed~\cite{Joyce:2016vqv,Gleyzes:2013ooa,
Bloomfield:2012ff,Sahni:1999gb,Sahni:2006pa,Tsujikawa:2013fta,
Peebles:2002gy,Baker:2012zs,Bamba:2012cp}: 
\begin{enumerate}
\item[$\bullet$] 
{\bf matter condensates or physical DE}, 
of which the simplest representatives are scalar condensates 
({\it quintessence} models);

\item[$\bullet$] 
{\bf modified gravity or geometrical DE}, 
which mimics dark energy 
by changing the relation between geometry and matter or 
by supplying additional geometric fields to general relativity.
\end{enumerate}
However, there is no impenetrable barrier between these two
possibilities, and more generically DE can be both physical and
geometrical, i.e. a new matter field has to be introduced and gravity
becomes modified, too. This just happens in the case of DE described 
by a non-minimally coupled scalar field considered in this paper. Recently 
the effective field theory (EFT) approach to dark 
energy~\cite{Gubitosi:2012hu,Bloomfield:2012ff,Gleyzes:2013ooa} 
has been developed. Its beauty is in that it presents a unified framework 
for both approaches, but its drawback is in that it does not immediately 
select the fundamental theory that lies behind some EFT. Nevertheless, 
different theories can be mapped onto the same class of EFTs, such that 
one can think of EFTs as identifying universality classes associated with 
DE models. 

The question of naturalness of initial conditions is not addressed in 
traditional approaches. For example, in quintessence models typically 
a quintessence field starts running from a value which is not a (local or 
global) minimum of the potential. Criticisms are often brushed away 
by noting that similar malady plagues most of inflationary models. 
Arguably the main benefit of this work is in that we construct a theory 
that naturally explains the initial field value -- which is accounted for
by the calculable amplitude of infrared field fluctuations during inflation
-- thus addressing this fundamental criticism. The program advocated 
here can be thought of as a third way for understanding DE, in that in 
our class of models a link is established between primordial inflation 
and dark energy. This link, among other things, can be  exploited when 
designing tests for these models.  

Observers have devoted a lot of effort (and observational time) to nail 
down as accurately as possible the amount (and distribution) of dark energy. 
Since its discovery in 1998~\cite{Riess:1998cb,Perlmutter:1998np}
a lot of progress has been made in improving the accuracy of DE 
measurements~\cite{Suzuki:2011hu,Parkinson:2012vd,
Palanque-Delabrouille:2014jca}. At this moment the Planck 
satelite~\cite{Ade:2015xua,Ade:2015rim} and the Dark Energy 
Survey (DES)~\cite{Abbott:2017wau} collaborations provide the 
most stringent bounds on dark energy. Presently $\Lambda$CDM, 
which assumes a cosmological constant equation of state $p\!=\!-\rho$ 
for dark energy, is consistent with all astronomical data.
Assuming a general constant equation of state parameter $p\!=\!w\rho$ 
(the so-called $w$CDM model), combined Planck and Type Ia supernovae 
data provides constraints~$w \!=\! -1.006\pm0.045$ (68\% CL). 

Considering dynamical dark energy models -- where $w$ that vary 
with redshift $z$ -- yields significantly relaxed constraints~\cite{Ade:2015rim}.
Still no statistically significant deviation of DE from an exact cosmological 
constant has been found~\footnote{Recent search for different channels 
of radioactive-type decay of DE with time-independent decay rates also 
resulted in upper limits on these rates less than the inverse present age 
of the Universe only \cite{Shafieloo:2016bpk}.}. The upcoming measurements 
of Large Synoptic  Survey Telescope (LSST)~\cite{Abate:2012za}, ESA's Euclid 
satellite mission~\cite{Amendola:2016saw} and the European Extreme 
Large Telescope (E-ELT)~\cite{Shearer:2010cn} will further constrain 
dynamical DE models. For example, the accuracy of 
Euclid~\cite{Amendola:2016saw} is projected to be a few  percent for 
simple (1 or 2 parameter models) and weaker for more involved models.
Furthermore, these missions will be able to test some prominent DE 
models, including clustering of DE, growth of Universe's structure, 
interactions of DE with itself and with other cosmological fluids,
and -- last but not least -- the class of models presented here.

However, if the present DE traces its origin to a very early stage of 
the Universe, initial conditions for its subsequent evolution were 
quantum. As a result, it can remain quantum even up to the present time.
Specifically in this work we investigate the influence of the quantum 
backreaction of a very light, nonminimally coupled spectator scalar 
field on the expansion dynamics of the Universe at late times. Our 
interest is to investigate what is the effect of inflationary quantum 
fluctuations that survive until late times and become comparable 
to -- and eventually dominate over -- the background nonrelativistic
matter energy density driving the matter-dominated expansion.
This idea goes back a while ago to~\cite{Sahni:1998at,Sahni:1999aq} 
where it was noted that quantum fluctuations of a very light scalar 
could provide a CC-like contribution to the Friedmann equations. In 
recent years, similar ideas were examined~\cite{Ringeval:2010hf, 
Glavan:2013mra, Aoki:2014ita, Glavan:2014uga, Aoki:2014dqa, 
Glavan:2015cut}. Most recently in~\cite{Glavan:2015cut} 
it was shown that, for certain ranges of model parameters, the 
backreaction remains small throughout the expansion of the Universe, 
and becomes relevant only at late times in matter era where it 
behaves approximately like a cosmological constant. Three parameters 
were introduced in this model: the total number of e-foldings of
inflation $N_I$, the nonminimal coupling $\xi$, and the scalar field 
mass $m$. The conditions these parameters need to meet for the 
scenario to unfold are~\cite{Glavan:2015cut}
\begin{equation}
\Bigl( \frac{m}{H_{DE}} \Bigr)^2\ll1 \, ,
\qquad
N_I < \frac{1}{8|\xi|} \ln
	\biggl[ 4\pi \Bigl( \frac{M_P}{H_I} \Bigr)^2 \biggr] \, ,
\qquad
0>\xi> -\frac{1}{6} \Bigl( \frac{m}{H_{DE}} \Bigr)^2 \, .
\end{equation}
The first condition guarantees that the scalar field stays light 
throughout the expansion, the second one comes from requiring 
the backreaction to remain small (perturbative) during inflation, 
and the third one ensures that the leading term in the backreaction 
at very late times is of the cosmological constant type. Here $H_I$ 
is the inflationary Hubble rate, $H_0$ the Hubble rate today, and 
$H_{DE}$ the Hubble rate at the onset of dark energy domination. 
The magnitude of the cosmological constant today implies another 
relation that determines the total number of e-folding of inflation 
$N_I$ in terms of the remaining two parameters,
\begin{equation}
N_I = \frac{1}{8|\xi|}  \ln \biggl[ 
	24\pi|\xi| \Bigl( \frac{M_P}{H_I} \Bigr)^2
	\Bigl( \frac{H_{DE}}{m} \Bigr)^2 
	\biggr] \, .
\label{intricate relation}
\end{equation}
It turns out this limits the range of the nonminimal coupling to
$0>\xi\ge-10^{-2}$, and the number of total e-foldings of inflation 
is $N_I\gtrsim 10^3$.

In the regime where the backreaction becomes comparable 
to the background it can no longer be treated perturbatively, 
but rather its effects have to be taken into account properly 
by solving the semiclassical Friedmann equations self-consistently, 
with the quantum backreaction as a source. In general, these are 
rather complicated integro-differential equations. In principle these 
equations can be solved numerically by discretizing both the time 
evolution and the integral over the modes for the backreaction.
An example of such computation was done in the context of 
reheating in~\cite{Suen:1987gu}, but it still presents a challenging 
numerical problem. Instead of the numerical approach we make 
use of the observation from the analytic 
computations~\cite{Glavan:2015cut,Glavan:2013mra,Glavan:2014uga}
that the quantum backreaction for very light fields is dominated 
by the infrared (IR) modes not only during inflation, but
throughout the history, and that the spectrum of these modes is 
inherited from the inflation era. The evolution of IR modes (which 
are largely amplified) during inflation (for small enough nonminimal 
coupling) is very accurately described by the formalism of stochastic 
inflation fully formulated in~\cite{Starobinsky:1986fx} (though
some of its simplest applications were already used 
in~\cite{Starobinsky:1982ee}). Here we adapt this method so that it 
becomes applicable to the evolution of IR modes in subsequent 
radiation and matter dominated eras of the Universe, and reproduce 
all the perturbative results of~\cite{Glavan:2015cut}. 
Such an approach to the problem was already advocated long ago 
in~\cite{Sahni:1999gb}, just after the original inception of the 
idea~\cite{Sahni:1998at,Sahni:1999aq}, but -- to our knowledge 
-- it has not been carried out. We then use this method to approximate 
the semiclassical Friedmann equations in the regime where quantum 
backreaction is very large and write them as a closed set of differential 
equations which are much simpler to solve than the original 
integro-differential equations of semi-classical gravity. This 
approximation captures accurately the backreaction effects. These 
equations are then solved numerically. We find that the backreaction 
indeed accelerates the Universe, driving it towards a late de Sitter phase, 
very much like the observed behaviour of the Universe today,
representing thus a novel dark energy candidate.

This paper is organized as follows. Section II presents the definition of
the scalar field model and the standard cosmological history. 
In section III the stochastic formalism is derived for the model at hand,
and in section IV the stochastic equations are solved in the regime of
small backreaction and the results of~\cite{Glavan:2015cut} are 
reproduced. Section V presents the numerical solution of the full 
semiclassical Friedmann equations at late times, when backreaction 
is no longer small. In section VI we summarize and discuss the principal 
results.


\section{Scalar field in FLRW}

In this section we define the cosmological background with the standard
expansion history, on which our spectator scalar field model is defined 
and quantized.


\subsection{FLRW background}

The metric of the $D$-dimensional spatially flat 
Friedman-Lema\^itre-Robertson-Walker 
(FLRW) spacetime is given by
\begin{equation}
g_{\mu\nu}=\text{diag}(-1,a^2(t),\dots,a^2(t))
\end{equation}
where~$a$ is the scale factor, and the derivative of its logarithmic 
with respect to time defines the Hubble expansion rate,~$H\!=\!\dot{a}/a$. 
The conventions that we use for the geometric quantities are
$\Gamma^{\alpha}_{\mu\nu} 
= \frac{1}{2} g^{\alpha\beta}(\partial_\mu g_{\nu\beta} 
+ \partial_\nu g_{\mu\beta} - \partial_\beta g_{\mu\nu})$
for the Christoffel symbol, 
$R^{\alpha}_{\mu\beta\nu} = \partial_\beta \Gamma^\alpha_{\mu\nu}
- \partial_\nu \Gamma^\alpha_{\beta\mu} 
+ \Gamma^{\rho}_{\mu\nu} \Gamma^\alpha_{\rho\beta}
- \Gamma^{\rho}_{\mu\beta} \Gamma^\alpha_{\rho\nu}$ 
for the Riemann tensor, and
$R_{\mu\nu} \!=\! {R^{\alpha}}_{\mu\alpha\nu}$ for the Ricci tensor.
A convenient quantity we often use instead of time is the number
of e-foldings~$N$, which for the period between 
some~$t_0$ and~$t$ is equal to~$N\!=\!\ln(a/a_0)$.

The classical evolution of a spatially homogeneous, spatially 
flat universe is governed by the Friedmann equations,
\begin{equation}
H^2 = \frac{2}{(D\!-\!1) (D\!-\!2) M_{\rm P}^2} \, \rho_c \, ,
\qquad
\dot{H} = - \frac{1}{(D\!-\!2) M_{\rm P}^2} \, (\rho_c + p_c) \, ,
\label{Friedman eqs}
\end{equation}
where 
$M_{\rm P} \!=\! (8\pi G)^{-1/2} \!\approx\! 2.45 \!\times\! 10^{18}{\,\rm GeV}$ 
is the reduced Planck mass~\footnote{In this paper we work in natural units, 
in which the Planck constant, $\hbar=1$ and the speed of light, $c=1$.}, $G$ 
is the Newton's constant, and $\rho_c\!=\!\rho_c(t)$ and $p_c \!=\! p_c(t)$ are 
the classical pressure and energy density of the (dominant) cosmological fluid, 
respectively. Here we are primarily interested in solving for the Universe's dynamics 
which also includes the quantum backreaction from inflationary quantum 
fluctuations of matter fields at the one-loop level. At the background level 
this approximation corresponds to what is known in the literature as 
{\it semiclassical gravity}, but the one-loop approximations goes beyond 
it since it includes higher order moments as well, so that background becomes 
stochastic effectively.

\begin{figure}
\includegraphics[width=7.8cm]{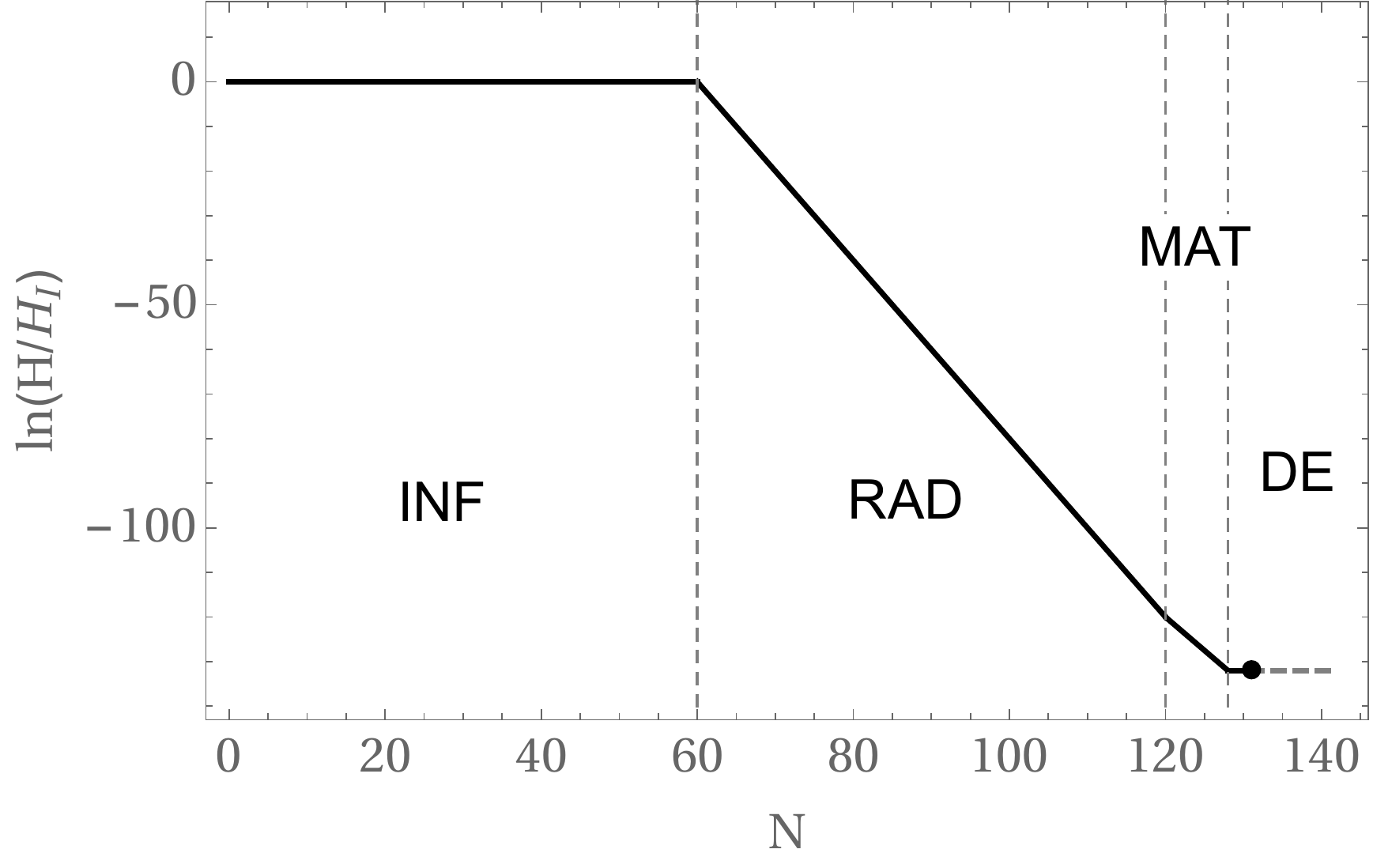} \ \ \ \ 
\includegraphics[width=8cm]{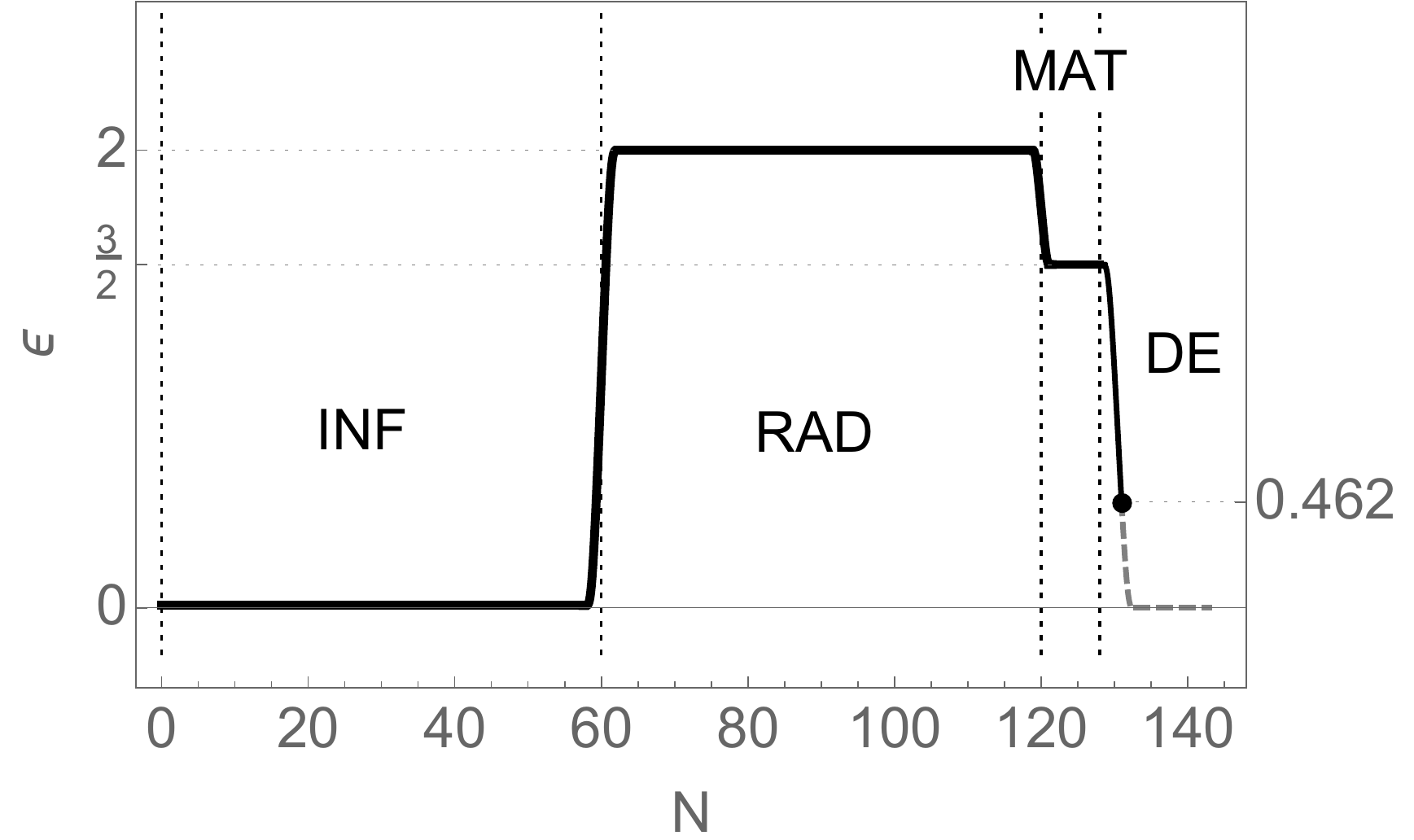}
\vskip-0.3cm
\caption{Schematic evolution of the Hubble parameter (left), and the $\epsilon$
parameter (right) during the history of the Universe. The dot on the curves designates 
where we are today.}
\label{history}
\end{figure}

Large portions of the expansion history of the Universe are dominated by 
a single cosmological fluid with a constant equation of state. These periods
are characterized by constant deceleration, or by a constant 
parameter~$\epsilon\!=\! -\dot{H}/H^2$, where~$\epsilon\!\approx\!0$
during inflation (where it is known as the slow-roll 
parameter),~$\epsilon\!=\!2$ during radiation-dominated period,
and~$\epsilon\!=\!3/2$ during matter dominated period (see Fig.~\ref{history}
for a schematic depiction of the evolution of~$\epsilon$ parameter).

During~$\epsilon\!=\!\text{const.}$ periods the scale factor and the Hubble
rate evolve as
\begin{equation}
a(t) = a_0 \bigl[ 
	1 + \epsilon H_0 (t\!-\!t_0) \bigr]^{\frac{1}{\epsilon}} \, ,
\qquad
H(t) = \frac{H_0}{1 + \epsilon H_0 (t\!-\!t_0)} \, ,
\end{equation}
where~$a_0$ and~$H_0$ refer to the values at some particular time
$t_0$, conveniently taken to be the beginning of a given 
constant~$\epsilon$ period. The evolution of the Hubble rate is
depicted schematically in Fig.~\ref{history}.

\subsection{Nonminimally coupled massive scalar}

The action for the nonminimally coupled scalar field in a curved 
background is
\begin{equation}
S[\Phi] = \int\! d^Dx \, \mathcal{L}_\Phi = \int\! d^D\!x \, \sqrt{-g} \, 
	\biggl\{ -\frac{1}{2} g^{\mu\nu} \partial_\mu \Phi \partial_\nu \Phi
		- \frac{1}{2}m^2\Phi^2 - \frac{1}{2}\xi R \Phi^2 \biggr\} \, ,
\label{the action}
\end{equation}
where $m$ is the scalar field mass,~$R$ is the Ricci scalar, 
and~$\xi$ is the nonminimal coupling with the sign convention 
where~$\xi\!=\xi_c\!=(D\!-\!2)/[4(D\!-\!1)]\!
\xrightarrow{D\rightarrow4}\!{1}/{6}$ corresponds to the conformal 
coupling. We consider this model on an FLRW background.

In order to quantize the scalar field we first need to define the 
canonical formalism. The canonical momentum is
\begin{equation}
\Pi(x) = \frac{\partial \mathcal{L}_\Phi}{\partial \dot{\Phi}(x)} =
	a^{D-1} \dot{\Phi}(x) \ , 
\end{equation}
and accordingly the Hamiltonian,
\begin{eqnarray}
\text{H}[\Phi;t) & = & \int\! d^{D-1}x\, 
	\Bigl[ \Pi(t,\vec{x}) \dot{\Phi}(t,\vec{x}) 
	- \mathcal{L}_\Phi (t,\vec{x}) \Bigr]
\nonumber \\
& = & \int\! d^{D-1}x\,  \biggl\{ \frac{a^{1-D}}{2} \Pi^2
	+ \frac{a^{D-3}}{2} \bigl( \vec{\nabla}\Phi \bigr)^{\!2}
	+ \frac{a^{D-1}}{2} 
		\Bigl[ m^2 + \xi(D\!-\!1)(D\!-\!2\epsilon)H^2 \Bigr] 
		\Phi^2 \biggr\} \, .
\end{eqnarray}
Canonical quantization is now accomplished by promoting 
the field and its conjugate momentum to operators, and their 
Poisson brackets to equal-time commutation relations,
\begin{equation}
\bigl[ \hat{\Phi}(t,\vec{x}), \hat{\Pi}(t,\vec{x}^{\,\prime}) \bigr]
	= i \delta^{D-1}(\vec{x}\!-\!\vec{x}^{\,\prime}) \, ,
\qquad
 \bigl[ \hat{\Phi}(t,\vec{x}), \hat{\Phi}(t,\vec{x}^{\,\prime}) \bigr] =0
	= \bigl[ \hat{\Pi}(t,\vec{x}), \hat{\Pi}(t,\vec{x}^{\,\prime}) \bigr] \, .
\end{equation}
The Heisenberg equations of motion for the field operators are now,
\begin{eqnarray}
\frac{d}{dt} \hat{\Phi}(t,\vec{x}) - a^{1-D} \, \hat{\Pi}(t,\vec{x}) = 0 \, ,
\label{FullEOM1}
\\
a^{1-D}\, \frac{d}{dt} \hat{\Pi}(t,\vec{x})
	- \frac{\nabla^2}{a^2} \hat{\Phi}(t,\vec{x})
	+ M^2(t)\, \hat{\Phi}(t,\vec{x}) = 0 \, ,
\label{FullEOM2}
\end{eqnarray}
where the effective mass-squared is
\begin{equation}
M^2(t) = m^2 + \xi R(t) = m^2 + \xi(D\!-\!1)(D\!-\!2\epsilon) H^2 \, .
\end{equation}
It is convenient to expand the field operators in Fourier modes,
\begin{eqnarray}
\hat{\Phi}(t,\vec{x}) & = &
	\mathlarger{\int}\!\! \frac{d^{D-1}k}{(2\pi)^{\frac{D-1}{2}}}
	\biggl\{ e^{i\vec{k}\cdot\vec{x}} \varphi(t,k) \, 
			\hat{b}(\vec{k}) 
		+ e^{-i\vec{k}\cdot\vec{x}} \varphi^*(t,k) \, 
			\hat{b}^{\dag}(\vec{k})  \biggr\} \, ,
\\
\hat{\Pi}(t,\vec{x}) & = &
	\mathlarger{\int}\!\! \frac{d^{D-1}k}{(2\pi)^{\frac{D-1}{2}}}
	\biggl\{ e^{i\vec{k}\cdot\vec{x}} a^{D-1} \dot{\varphi}(t,k) \, 
			\hat{b}(\vec{k}) 
		+ e^{-i\vec{k}\cdot\vec{x}} a^{D-1} \dot{\varphi}^*(t,k) \, 
			\hat{b}^{\dag}(\vec{k})  \biggr\} \, ,
\end{eqnarray}
where the expansion of the conjugate momentum operator is such 
that it automatically satisfies Eq.~(\ref{FullEOM1}). Observations 
tell us that the Universe is nearly spatially homogeneous on large scales. 
This motivates a spatially homogeneous {\it Ansatz} for the mode 
function $\varphi(t,k)$,  where $k$  denotes the modulus of the
comoving momentum $k\!=\!\|\vec{k}\|$, and the annihilation and 
creation operators, $\hat{b}$ and $\hat{b}^{\dag}$, satisfy canonical 
commutation relations
\begin{equation}
\bigl[ \hat{b}(\vec{k}) , \hat{b}^{\dag}(\vec{k}^{\,\prime}) \bigr]
	= \delta^{D-1}(\vec{k}\!-\!\vec{k}^{\,\prime}) \, , 
\qquad
\bigl[ \hat{b}(\vec{k}) , \hat{b}(\vec{k}^{\prime}) \bigr] = 0
	= \bigl[ \hat{b}^{\dag} (\vec{k}) , 
		\hat{b}^{\dag}(\vec{k}^{\prime}) \bigr] \, .
\end{equation}
This puts a constraint on the mode function in the form of a Wronskian 
normalization,
\begin{equation}
\varphi(t,k) \dot{\varphi}^*(t,k)
	- \dot{\varphi}(t,k) \varphi^*(t,k) = i \, a^{1-D} \, .
\end{equation}
The equation of motion for  the mode function
inferred from~(\ref{FullEOM1}-\ref{FullEOM2}) is
\begin{equation}
\ddot{\varphi}(t,k) + (D\!-\!1)H\dot{\varphi}(t,k)
	+ \frac{k^2}{a^2}\varphi(t,k) + M^2(t)\, \varphi(t,k) = 0 \, .
\label{modeEOM}
\end{equation}
The Hilbert space of states is constructed in the standard manner 
by defining a vacuum state $|\Omega\rangle$, which is annihilated 
by all the annihilation operators, 
$\hat b(\vec{k})|\Omega\rangle\!=\!0, \forall \vec{k}$. Then the rest of 
the state space is constructed by creation operators 
acting on $|\Omega\rangle$.

The energy-momentum tensor operator of the scalar field is given by
\begin{eqnarray}
\hat{T}_{\mu\nu} & = & 
	\frac{-2}{\sqrt{-g}} \frac{\delta S[\Phi]}{\delta g^{\mu\nu}}
	\Bigr|_{\Phi\rightarrow \hat{\Phi}}
\nonumber \\
	& = & \partial_\mu \hat{\Phi} \, \partial_\nu\hat{\Phi}
		- \frac{1}{2} g_{\mu\nu} g^{\alpha\beta} 
			\partial_\alpha \hat{\Phi} \, \partial_\beta \hat{\Phi}
		- \frac{m^2}{2} g_{\mu\nu} \hat{\Phi}^2
		+ \xi \bigl[ G_{\mu\nu}  - \nabla_\mu \nabla_\nu 
			+ g_{\mu\nu} \square \bigr]
			\hat{\Phi}^2 \, .
\end{eqnarray}
Its expectation value $\langle\cdot\rangle$ with respect to a 
homogeneous and isotropic state $|\Omega\rangle$
defined on a FLRW background (\ref{Friedman eqs}) 
is diagonal and takes the form of ideal fluid,
\begin{eqnarray}
\rho_Q & = & - \langle \hat{T}^0{}_0\rangle =
	\frac{1}{2} \langle ( \partial_0 \hat{\Phi} )^2 \rangle 
	+ \frac{1}{2a^2} \langle (\vec{\nabla}\Phi)^2 \rangle
	+ \frac{m^2}{2} \langle \hat{\Phi}^2 \rangle 
\nonumber \\
&&	\hspace{3cm}
	+ \xi \Bigl[ \frac{1}{2} (D\!-\!1)(D\!-\!2) H^2
		+ (D\!-\!1) H \partial_0 \Bigr] 
		\langle \hat{\Phi}^2 \rangle \, ,
\\
p_Q \delta^i_j & = & \langle \hat{T}^i{}_j \rangle
	=  \delta^i_j \biggl\{ \frac{1}{2} 
		\langle (\partial_0\hat{\Phi})^2 \rangle 
	- \frac{(D\!-\!3)}{2(D\!-\!1)a^2} 
		\langle (\vec{\nabla} \hat{\Phi})^2 \rangle
	- \frac{m^2}{2} \langle \hat{\Phi}^2 \rangle 
\nonumber \\
&&	\hspace{2.5cm}
	- \xi \Bigl[ \frac{1}{2} (D\!-\!2) 
		\bigl( 2\dot{H} + (D\!-\!1) H^2 \bigr)
		+ \partial_0^2 + (D\!-\!2) H \partial_0 \Bigr] 
		\langle \hat{\Phi}^2 \rangle \biggr\} \, ,
\end{eqnarray}
where $\rho_Q \!=\! \rho_Q(t)$ and  $p_Q \!=\! p_Q(t)$ 
are spatially homogeneous.


\section{Stochastic formalism}
\label{sec: Stochastic formalism}

Here we briefly introduce the stochastic formalism
of~\cite{Starobinsky:1986fx} for scalar fields in expanding 
cosmological space. The dominant contribution to the light 
scalar field correlators in inflation comes from the superhorizon
modes (of wavelengths $k\!>\!1/aH$). This is also true for the very 
light or massless scalar fields in subsequent evolution of the Universe, 
namely during radiation and matter periods~\cite{Glavan:2015cut,
Glavan:2014uga}. The slow-roll approximation usually employed to 
derive the stochastic equations need not be correct in situations 
when the scalar becomes very massive, meaning~$m\!\gtrsim \!H$.
Since during decelerating periods of expansion the Hubble rate 
decays, this can eventually become true. This is why we do not use 
the slow-roll approximation in the stochastic formalism, but rather 
derive the equations for all three IR correlators. These reduce to the 
standard equations when the slow-roll hierarchy between  the
correlators is present.


\subsection{Equations of motion for field operators}

We split the field operators into the contributions from the long 
wavelength modes and from the short wavelength modes as follows,
\begin{equation}
\hat{\Phi}(t,\vec{x}) = \hat{\phi}(t,\vec{x}) 
	+ \hat{\phi}_{\text{s}}(t,\vec{x}) \, ,
\qquad
\hat{\Pi}(t,\vec{x}) = \hat{\pi}(t,\vec{x}) 
	+ \hat{\pi}_{\text{s}}(t,\vec{x}) \, ,
\label{split}
\end{equation}
separated by a comoving scale $\mu a H$, where $0<\mu\!\ll\!1$ is 
the control parameter of the splitting. This splitting is most 
conveniently imposed in Fourier space where the long wavelength 
parts are
\begin{eqnarray}
\hat{\phi}(t,\vec{x}) & = & 
	\mathlarger{\int}\!\! \frac{d^{D-1}k}{(2\pi)^{\frac{D-1}{2}}} \,
	\theta\bigl( \mu a H \!-\! \| \vec{k} \| \bigr)
	\biggl\{ e^{i\vec{k}\cdot\vec{x}} \varphi(t,k) \, 
			\hat{b}(\vec{k}) 
		+ e^{-i\vec{k}\cdot\vec{x}} \varphi^*(t,k) \, 
			\hat{b}^{\dag}(\vec{k})  \biggr\} \, ,
\label{PhiLong}
\\
\hat{\pi}(t,\vec{x}) & = &
	\mathlarger{\int}\!\! \frac{d^{D-1}k}{(2\pi)^{\frac{D-1}{2}}} \,
	\theta\bigl( \mu a H \!-\! \| \vec{k} \| \bigr)
	\biggl\{ e^{i\vec{k}\cdot\vec{x}} a^{D-1} \dot{\varphi}(t,k) \, 
			\hat{b}(\vec{k}) 
		+ e^{-i\vec{k}\cdot\vec{x}} a^{D-1} \dot{\varphi}^*(t,k) \, 
			\hat{b}^{\dag}(\vec{k})  \biggr\} \, , \qquad
\label{PiLong}
\end{eqnarray}
and the short wavelength parts
\begin{eqnarray}
\hat{\phi}_{\text{s}}(t,\vec{x}) & = &
	\mathlarger{\int}\!\! \frac{d^{D-1}k}{(2\pi)^{\frac{D-1}{2}}} \,
	\theta\bigl( \| \vec{k} \| \!-\!\mu a H \bigr)
	\biggl\{ e^{i\vec{k}\cdot\vec{x}} \varphi(t,k) \, 
			\hat{b}(\vec{k}) 
		+ e^{-i\vec{k}\cdot\vec{x}} \varphi^*(t,k) \, 
			\hat{b}^{\dag}(\vec{k})  \biggr\} \, ,
\label{PhiShort}
\\
\hat{\pi}_{\text{s}}(t,\vec{x}) & = &
	\mathlarger{\int}\!\! \frac{d^{D-1}k}{(2\pi)^{\frac{D-1}{2}}} \,
	\theta\bigl( \| \vec{k} \| \!-\!\mu a H \bigr)
	\biggl\{ e^{i\vec{k}\cdot\vec{x}} a^{D-1} \dot{\varphi}(t,k) \, 
			\hat{b}(\vec{k}) 
		+ e^{-i\vec{k}\cdot\vec{x}} a^{D-1} \dot{\varphi}^*(t,k) \, 
			\hat{b}^{\dag}(\vec{k})  \biggr\} \, , \qquad
\label{PiShort}
\end{eqnarray}
where, for simplicity, we took the Heaviside $\theta$-function for 
the window functions. We want to derive the analog of equations 
of motion \eqref{FullEOM1} and \eqref{FullEOM2}, but just for the 
long wavelength parts of the field operators \eqref{PhiLong} 
and \eqref{PiLong}. Making use of the equation of 
motion~(\ref{modeEOM}) for the mode function the resulting equations are
\begin{eqnarray}
\frac{d}{dt} \hat{\phi}(t,\boldsymbol{x}) 
	- a^{1-D}\, \hat{\pi}(t,\boldsymbol{x})
	= \hat{f}_\phi(t,\boldsymbol{x}) \, ,
\label{EOMdPhiLong}
\\
a^{1-D}\, \frac{d}{dt} \hat{\pi}(t,\boldsymbol{x})
	- \frac{\nabla^2}{a^2} \hat{\phi}(t,\boldsymbol{x})
	+ M^2(t)\, \hat{\phi}(t,\boldsymbol{x})
	= a^{1-D} \hat{f}_\pi (t,\boldsymbol{x}) \, ,
\label{EOMdPiLong}
\end{eqnarray}
where the sources -- which originate from the coupling between 
the short and long wavelength fields 
-- are
\begin{eqnarray}
\hat{f}_\phi(t,\vec{x})
	& = & \mu a H^2(1\!-\!\epsilon) 
	\mathlarger{\int}\!\! \frac{d^{D-1}k}{(2\pi)^{\frac{D-1}{2}}} \,
	\delta\bigl( \|\vec{k}\|\!-\!\mu a H \bigr)
	\biggl\{ e^{i\vec{k}\cdot\vec{x}} \varphi(t,k) \, 
			\hat{b}(\vec{k}) 
		+ e^{-i\vec{k}\cdot\vec{x}} \varphi^*(t,k) \, 
			\hat{b}^{\dag}(\vec{k})  \biggr\} \, .
\label{fPhi}
\\
\hat{f}_\pi(t,\vec{x})
	& = & \mu a^D H^2(1\!-\!\epsilon) 
	\mathlarger{\int}\!\! \frac{d^{D-1}k}{(2\pi)^{\frac{D-1}{2}}} \,
	\delta\bigl( \|\vec{k}\|\!-\!\mu a H \bigr)
	\biggl\{ e^{i\vec{k}\cdot\vec{x}} \dot{\varphi}(t,k) \, 
			\hat{b}(\vec{k}) 
		+ e^{-i\vec{k}\cdot\vec{x}} \dot{\varphi}^*(t,k) \, 
			\hat{b}^{\dag}(\vec{k})  \biggr\} \, . \qquad
\label{fPi}
\end{eqnarray}
The two sources \eqref{fPhi} and \eqref{fPi} are to be considered 
as stochastic forces acting on the long wavelength fields, and here 
they originate from the effect of modes leaving (entering) the Hubble 
sphere due to accelerating (decelerating) expansion.


\subsection{Equations of motion for IR correlators}

Here we derive the equations for motion for the coincident 
long wavelength (infrared, IR) two-point functions, which we 
define to be conveniently rescaled,
\begin{eqnarray}
\Delta_{\phi\phi}(t) & \equiv & 
	\Bigl\langle \hat{\phi}^2 (t,\vec{x}) \Bigr\rangle \, ,
\label{DeltaPhiPhi}
\\
\Delta_{\phi\pi}(t) & \equiv & 
	\frac{1}{a^3(t) H(t)}\Bigl\langle \bigl\{ \hat{\phi}(t,\vec{x}) ,
						\hat{\pi}(t,\vec{x}) \bigr\} \Bigr\rangle \, ,
\label{DeltaPhiPi}
\\
\Delta_{\pi\pi}(t) & \equiv & 
	\frac{1}{a^6(t) H^2(t)} \Bigl\langle \hat{\pi}^2 (t,\vec{x}) 
	\Bigr\rangle \, .
\label{DeltaPiPi}
\end{eqnarray}
The convenience of time dependent factors in the definitions 
above is in that all of these correlators become of the same 
dimension and the way they appear in the equations allows us 
to compare them  directly which makes comparing their
magnitudes straightforward. From now on we set $D\!=\!4$ 
since the UV divergences are captured by the short wavelength 
part of the fields. The equations of motion for these correlators 
follow from the equations of motion~\eqref{EOMdPhiLong} 
and~\eqref{EOMdPiLong} for the long wavelength parts of the 
field operators (in $D\!=\!4$),
\begin{eqnarray}
\frac{d}{dN} \Delta_{\phi\phi} - \Delta_{\phi\pi} = n_{\phi\phi} \, ,
\label{eomPhiPhi}
\\
\frac{d}{dN} \Delta_{\phi\pi} + (3-\epsilon) \Delta_{\phi\pi}
	- 2 \Delta_{\pi\pi} + 2 \Bigl( \frac{M}{H} \Bigr)^{\!2} 
		\Delta_{\phi\phi} = n_{\phi\pi} \, ,
\label{eomPhiPi}
\\
\frac{d}{dN} \Delta_{\pi\pi} + 2(3 - \epsilon) \Delta_{\pi\pi}
	+ \Bigl( \frac{M}{H} \Bigr)^{\!2} \Delta_{\phi\pi} = n_{\pi\pi} \, ,
\label{eomPiPi}
\end{eqnarray}
In deriving these equations we have thrown away gradients, in 
particular terms of the form $\langle (\boldsymbol{\nabla}\phi)^2 \rangle$, 
since they are suppressed by a factor of $\mu^2\!\ll\!1$ compared to 
the rest. This allows one to close the set of equations. Also we have 
switched to the number of e-foldings $N\!=\! \ln(a)$ as the time
variable (with the choice of time (gauge) that corresponds to
$a\!=\!1$ at the beginning of inflation), and
$\epsilon \!=\! -\dot{H}/H^2$. The stochastic sources on the right hand 
side of~(\ref{eomPhiPhi}--\ref{eomPiPi}) are the coincident field-noise 
correlators,
\begin{eqnarray}
n_{\phi\phi} & \equiv & \frac{1}{H(t)} \Bigl\langle 
	\bigl\{ \hat{f}_{\phi}(t,\boldsymbol{x}) , 
		\hat{\phi}(t,\boldsymbol{x}) \bigr\} 
		\Big\rangle \, ,
\label{n phiphi}\\
n_{\phi\pi} & \equiv & \frac{1}{a^3(t)H^2(t)} \biggl[ 
	\Bigl\langle
	\bigl\{ \hat{f}_{\phi}(t,\boldsymbol{x}) , 
		\hat{\pi}(t,\boldsymbol{x}) \bigr\} \Big\rangle +
	\bigl\{ \hat{f}_{\pi}(t,\boldsymbol{x}) , 
		\hat{\phi}(t,\boldsymbol{x}) \bigr\} \Big\rangle
		\biggr] \, ,
\label{n phipi}
\\
n_{\pi\pi} & \equiv & \frac{1}{a^6(t)H^3(t)} \Bigl\langle 
	\bigl\{ \hat{f}_{\pi}(t,\boldsymbol{x}) , 
		\hat{\pi}(t,\boldsymbol{x}) \bigr\} 
		\Big\rangle \, .
\label{n pipi}
\end{eqnarray}
These are straightforwardly computed 
from~(\ref{PhiLong}--\ref{PiLong}) and~(\ref{fPhi}--\ref{fPi}) 
to be
\begin{eqnarray}
\Bigl\langle \bigl\{ \hat{f}_\phi(t,\vec{x}), 
	\hat{\phi}(t,\vec{x}) \bigr\} \Bigr\rangle
	& = & \frac{1}{2\pi^2} \, \mu^3 a^3H^4 (1\!-\!\epsilon)
		\Bigl[ |\varphi(t,k)|^2 \Bigr]_{k=\mu a H} \, ,
\label{CORRfPhiPhi}
\\
\Bigl\langle \bigl\{ \hat{f}_\phi(t,\vec{x}), 
	\hat{\pi}(t,\vec{x}) \bigr\} \Bigr\rangle
	& = & \frac{1}{4\pi^2} \, \mu^3 a^6H^4 (1\!-\!\epsilon)
		\biggl[ \frac{\partial}{\partial t}|\varphi(t,k)|^2 
			\biggr]_{k=\mu a H} \, ,
\label{CORRfPhiPi}
\\
\Bigl\langle \bigl\{ \hat{f}_\pi(t,\vec{x}), 
	\hat{\phi}(t,\vec{x}) \bigr\} \Bigr\rangle
	& = & \frac{1}{4\pi^2} \, \mu^3 a^6H^4 (1\!-\!\epsilon)
		\biggl[ \frac{\partial}{\partial t}|\varphi(t,k)|^2 
			\biggr]_{k=\mu a H} \, ,
\label{CORRfPiPhi}
\\
\Bigl\langle \bigl\{ \hat{f}_\pi(t,\vec{x}), 
	\hat{\pi}(t,\vec{x}) \bigr\} \Bigr\rangle
	& = & \frac{1}{2\pi^2} \, \mu^3 a^9H^4 (1\!-\!\epsilon)
		\Bigl[ |\dot{\varphi}(t,k)|^2 \Bigr]_{k=\mu a H} \, ,
\label{CORRfPiPi}
\end{eqnarray}
where the anticommutator is defined 
as~$\{ \hat{A}, \hat{B} \} \!\equiv\! \hat{A}\hat{B} \!+\! \hat{B}\hat{A}$,
and where we have used that $\theta(0) \!=\! {1}/{2}$.~\footnote{ 
In the calculation there appears a product 
$\theta(x)\delta(x)\!\rightarrow\! (1/2)\delta(x)$,
which can be justified by using a limiting procedure on a 
smooth window function.}  The sources~(\ref{n phiphi}-\ref{n pipi})
are referred to as the stochastic sources in analogy with the stochastic
inflation formalism, though their physical interpretation during decelerating
cosmological eras is not necessarily the one of standard stochastic sources.
During decelerating eras ($\epsilon\!>\!1$) these sources have a 
negative overall sign, as evident from~(\ref{CORRfPhiPhi}-\ref{CORRfPiPi}),
and in fact contribute towards decreasing the correlators. This is not
at all surprising if one notes that these sources arise as a time-dependent split
of the modes into super-Hubble and sub-Hubble ones. During decelerated
eras modes {\it leak} from the super-Hubble to sub-Hubble phase space
(as opposed to accelerating periods where the leak is reversed) which
accounts for the decrease in the IR correlators.

The energy density and pressure expectation values can be 
expressed in terms of the coincident IR correlators,
\begin{eqnarray}
\rho_Q & \approx & \frac{H^2}{2} \biggl\{ \Delta_{\pi\pi} 
	+ 6\xi  \Delta_{\phi\pi} + \Bigl[ 6\xi + \Bigl( \frac{m}{H} \Bigr)^{\!2} 
		\Bigr] \Delta_{\phi\phi}  \biggr\} \, ,
\label{backreactionRHO}
\\
p_Q & \approx & \frac{H^2}{2} \biggl\{ (1\!-\!4\xi) \Delta_{\pi\pi} 
	+ 2 \xi \Delta_{\phi\pi} + \Bigl[ - 2\xi (3 \!-\! 2\epsilon)
	+ 24\xi^2 (2 \!-\! \epsilon) 
		- \Bigl( \frac{m}{H} \Bigr)^{\!2}(1 \!-\! 4\xi) \Bigr] 
		\Delta_{\phi\phi} \biggr\} \, . \qquad
\label{backreactionP}
\end{eqnarray}
The gradients were thrown away here on the same grounds 
as in the equations of motion~(\ref{eomPhiPhi}-\ref{eomPiPi}) 
of the previous subsection. The short wavelength contributions 
were neglected since -- after renormalization -- except possibly 
during inflation, they contribute negligibly~\footnote{One can 
namely show that UV contributions to $\rho_Q$ and $p_Q$ 
are suppressed as $\sim H^4$.}. These expressions for energy 
density and pressure can now be used to quantum-correct the 
Friedmann equations (in the sense that one should exact the 
replacements, $\rho_c\rightarrow \rho_c+\rho_Q$ and 
$p_c\rightarrow p_c+p_Q$ in~(\ref{Friedman eqs}))
and together with equations of 
motion~(\ref{eomPhiPhi}-\ref{eomPiPi}) form a closed set of 
stochastic differential equations. It turns out that at late times 
(during radiation and matter era) stochastic sources contribute 
negligibly which will greatly simplify solving these equations.


\section{Comparison with field theoretic calculations}
\label{Comparison with field theoretic calculations}

Here we employ the stochastic formalism of the preceding 
section to calculate the backreaction of the scalar field quantum 
fluctuations during the three relevant cosmological periods and 
solve them perturbatively. These solutions are applicable in the 
regime where the quantum 
backreaction~(\ref{backreactionRHO}-\ref{backreactionP})
 is still negligible in comparison with the classical sources 
 $\rho_c$ and $p_c$ in~(\ref{Friedman eqs}). The sources for 
 the equations of motion~(\ref{eomPhiPhi}--\ref{eomPiPi}) 
are determined using the mode functions from~\cite{Glavan:2015cut}. 
This section serves to reaffirm the results of~\cite{Glavan:2015cut} 
as well as to test the (perturbative) correctness of the stochastic 
formalism.


\subsection{De Sitter inflationary period}

The results presented in this subsection were already derived in 
\cite{Finelli:2008zg,Finelli:2010sh}, both for test fields and for 
cosmological perturbations, and on the more general slow roll 
inflationary background.

The Chernikov-Tagirov-Bunch-Davies (CTBD) mode function of the 
scalar during the de Sitter inflation is
\begin{equation}
\varphi(t,k) = \sqrt{\frac{\pi}{4a^3H_I}} \ 
	H_{\nu_I}^{(1)} \Bigl( \frac{k}{aH_I} \Bigr) \, ,
\qquad
\nu_I = \sqrt{\tfrac{9}{4}-12\xi 
	- \bigl( \tfrac{m}{H_I} \bigr)^{\!2} } \, .
\label{mode function}
\end{equation}
where $H_\nu^{(1)}$ stands for the Hankel function of the first kind,
and $H_I\!=\!{\rm const.}$ is the inflationary Hubble rate. This mode 
function coincides with the one of the adiabatic vacuum state for 
the subhorizon modes. The leading order IR expansion of the mode 
function~(\ref{mode function}) (for~$k\!<\!\mu a H_I \!\ll\! a H_I$) is
\begin{equation}
\varphi(t,k) \approx \frac{(-i)}{\sqrt{\pi}}\, 
	2^{\nu_I-1} \Gamma(\nu_I) a^{\nu_I-3/2} H_I^{\nu_I-1/2} k^{-\nu_I} \, .
\end{equation}
With this we can determine stochastic 
sources~(\ref{n phiphi}--\ref{n pipi}) for the equations of motion,
\begin{equation}
n_{\phi\phi} =
	\frac{2^{2\nu_I-3}}{\pi^3} \, 
		\Gamma^2(\nu_I) \mu^{3-2\nu_I} H_I^2 \, ,
\quad
n_{\phi\pi} = 2 \bigl( \nu_I \!-\! \tfrac{3}{2} \bigr) n_{\phi\phi} \, ,
\quad
n_{\pi\pi} = \bigl( \nu_I \!-\! \tfrac{3}{2} \bigr)^2 n_{\phi\phi} \, .
\end{equation}
which are just constant.
To leading order in nonminimal coupling and 
mass,~$0 \!<\! |\xi|, m^2/H_I^2 \!\ll\! 1$, they are independent of 
the separation scale $\mu$,
\begin{equation}
n_{\phi\phi} = \frac{H_I^2}{4\pi^2} \, ,
\qquad
n_{\phi\pi} = -\frac{H_I^2}{6\pi^2} X \, ,
\qquad
n_{\pi\pi} = \frac{H_I^2}{32\pi^2} X^2 \, ,
\end{equation}
where the shorthand notation 
is~$X \!=\! 12\xi  +  (m/H_I)^2 $,
and it is clear that~$|X| \!\ll\! 1$, which can be used as an expansion 
parameter, encompassing both minimally coupled and massive case.
Parameter $X$ plays the role of an effective mass squared (in units 
of $H_I^2$) of the field during inflation. It can can take negative values 
$X<0$, signaling tachyonic particle production. This is not troubling, 
since the tachyonic nature of $X$ depends on the background spacetime, 
and is not present {\it e.g.} in flat space or in radiation 
era.~\footnote{The tachyonic nature of the IR modes during inflation 
is not problematic, since one can use {\it e.g.} pre-inflationary radiation 
era to regulate the infrared sector of the theory by a smooth matching 
of inflationary modes onto the radiation era modes~\cite{Janssen:2009nz}.}

The equations of motion for the coincident 
correlators~(\ref{eomPhiPhi}-\ref{eomPiPi}) are now,
\begin{eqnarray}
\frac{d}{dN} \Delta_{\phi\phi} - \Delta_{\phi\pi}
	& = & \frac{H_I^2}{4\pi^2} \, ,
\\
\frac{d}{dN} \Delta_{\phi\pi} + 3 \Delta_{\phi\pi}
	- 2 \Delta_{\pi\pi} + 2X  \Delta_{\phi\phi}
	& = & - \frac{H_I^2}{6\pi^2} X \, ,
\\
\frac{d}{dN} \Delta_{\pi\pi} + 6 \Delta_{\pi\pi}
	+ X \Delta_{\phi\pi} 
	& = & \frac{H_I^2}{36\pi^2} X^2 \, .
\end{eqnarray}
The natural initial conditions for these equations are
\begin{equation}
\Delta_{\phi\phi}(0) = 0 \,, \qquad
\Delta_{\phi\pi}(0) = 0 \,, \qquad
\Delta_{\pi\pi}(0) = 0 \, .
\label{inflation initial conditions}
\end{equation}
These can be seen as arising from the set of modes which were 
initially super-Hubble ($k<\mu H_I$), meaning that initially the IR phase 
space is zero (this point we identify with the beginning of inflation at 
which $a=1$). This is not strictly speaking so, we rather assume that the 
super-Hubble modes at the beginning of inflation are suppressed due to 
some physical mechanism (for examples of how this can be done in practice
see~\cite{Janssen:2009nz}). In practice it is sufficient to assume
that $\Delta_{\phi\phi} \!<\! H_I^2$ at the beginning of inflation. 
This results in them contributing subdominantly 
at late times.

Now the leading order solution in $|X|\!\ll\!1$ for these IR correlators is
\begin{equation}
\Delta_{\phi\phi}(N) = \frac{3H_I^2}{8\pi^2X} 
	\Bigl[ 1 - e^{-\frac{2}{3}XN} \Bigr] \, ,
\quad
\Delta_{\phi\pi}(N) = - \frac{2X}{3} \Delta_{\phi\phi}(N) \, ,
\quad
\Delta_{\pi\pi}(N) = \frac{X^2}{9} \Delta_{\phi\phi}(N) \, ,
\label{correlators inflation}
\end{equation}
The subdominant contributions are suppressed by additional factors 
of $X^3$ compared to the leading order 
one~\footnote{The exact form of the exponent in 
solutions~\eqref{correlators inflation}
is $\bigl[-3\bigl(1 \!-\! \sqrt{1 \!-\! 4X/9} \,\bigr)N\bigr]$, but we have 
expanded it under the assumption $|X|^2N_I \!\ll\! 1$, where $N_I$ 
is the total number of e-folding. This will be true in our case.}, and also 
decay much faster with the number of e-foldings. Given the initial 
conditions~\eqref{inflation initial conditions}, the stochastic sources 
determine the amplitude of the IR correlators. These sources also induce 
the hierarchy between the correlators which survives until very late times.

There are two interesting limits we discuss briefly in the two following 
subsections.


\subsubsection{Minimally coupled limit}

The minimally coupled limit corresponds to taking $X\!=\!m^2/H_I^2$.
There are two different cases to consider, the first one of which is
the ``long'' inflation limit $N_I \!\gg\! (H_I/m)^2$, where the correlators
eventually saturate to constant (de Sitter invariant) values,
\begin{equation}
\Delta_{\phi\phi} = \frac{3H_I^2}{8\pi^2} 
	\Bigl( \frac{m}{H_I} \Bigr)^{\!\!-2} \, ,
\qquad
\Delta_{\phi\pi} = 
	- \frac{2}{3} \Bigl( \frac{m}{H_I} \Bigr)^{\!2} \Delta_{\phi\phi} \, ,
\qquad
\Delta_{\pi\pi} = 
	\frac{1}{9} \Bigl( \frac{m}{H_I} \Bigr)^{\!4} \Delta_{\phi\phi}  \, .
\label{minimalEoI}
\end{equation}
The other case is the limit
of ``short'' inflation $N_I \! \ll \! (H_I/m)^2$, where the correlators 
do not have enough time to saturate, and their values at the end of 
inflation are linear in the total number of e-foldings of inflation,
\begin{equation}
\Delta_{\phi\phi}(N_I) = \frac{H_I^2}{4\pi^2} N_I \, , 
\qquad
\Delta_{\phi\pi}(N_I)= 
	- \frac{2}{3}\Bigl( \frac{m}{H_I} \Bigr)^{\!2} \Delta_{\phi\phi}(N_I) \, , 
\qquad
\Delta_{\pi\pi}(N_I) = 
	\frac{1}{9} \Bigl( \frac{m}{H_I} \Bigr)^{\!4} \Delta_{\phi\phi}(N_I) \, .
\label{minimal end of infl}
\end{equation}
In both cases the hierarchy between the correlators is dictated 
by the small mass.

In the case of ``long'' inflation the energy density and pressure by the
end of inflationary period are
\begin{equation}
\rho_Q = \frac{61 H_I^4}{960\pi^2} \, ,
\qquad
p_Q = -\rho_Q \, ,
\label{long inf}
\end{equation}
and in the case of ``short'' inflation they are
\begin{equation}
\rho_Q = - \frac{119 H_I^4}{960 \pi^2}
	+ \frac{H_I^2 m^2}{8\pi^2} N_I \, ,
\qquad
p_Q = -\rho_Q \, .
\end{equation}
Note that the contribution from the conformal anomaly
$\rho_{CA} \!=\! -p_{CA} \!=\! -119 H_I^4 /(960\pi^2)$ is included in both
expressions since it is not negligible in these cases. In subsequent periods
though, its contribution will be negligible.  It should be noted 
that for our purposes the ``long'' inflation limit is extremely long
(about $10^{13}$ efoldings), and the result~(\ref{long inf}) is computed
under the assumption of exact de Sitter inflation (exactly flat plateau 
inflation). In cases of small deviations from such assumptions on inflation 
the scalar spectator does not necessarily reach stationary values
for its correlators as pointed out in~\cite{Hardwick:2017fjo}, and one needs to 
be more careful when relating the correlator values of the spectator to
the inflationary period history. What is of primary interest for us in this work is 
to establish that the scalar can attain super-Planckian correlator 
values by the end of inflation.
The details of preciasely how it attains these values will introduce model dependent corrections to the 
relation~(\ref{intricate relation}) relating the number of e-foldings of inflation to the properties
of the late time Universe.


\subsubsection{Massless limit}

In the limit of negative nonminimal coupling which dominates over the 
mass term,~$1\!\gg\!|\xi|\!\gg\! (m/H_I)^2$, to leading order the field is
effectively massless, and the correlators grow exponentially with the
total number of e-foldings of inflation,
\begin{align}
\Delta_{\phi\phi}(N_I) 
	= \frac{H_I^2}{32\pi^2 |\xi|} \, e^{8 |\xi| N_I} \, , 
\quad
\Delta_{\phi\pi}(N_I) = 8|\xi| \,\Delta_{\phi\phi}(N_I) \, , 
\quad
\Delta_{\pi\pi}(N_I) = 16|\xi| \,\Delta_{\phi\phi}(N_I) \, .
\label{massless end of infl}
\end{align}
This growth of correlators during inflation is important for us for the 
model building of DE.

Note that in this case the leading order behavior of the correlators 
at late times in inflation is independent of the stochastic source, namely 
their growth is dominated by the instability of IR modes during inflation. 
Nevertheless, the stochastic sources, which in this case contribute
significantly only during the beginning of inflationary period, fix the 
amplitude and the hierarchy 
of the correlators, accounting for the fact that all the contributing modes 
are of UV origin.

The energy density and pressure of the backreaction by the end of 
inflationary period in this case are
\begin{equation}
\rho_Q =- \frac{3 H_I^4}{32\pi^2} \, e^{8 |\xi| N_I} \, ,
\qquad
p_Q = - \rho_Q \, .
\end{equation}
%


\subsection{Radiation period}

In radiation period we have $\epsilon_R \!=\! 2$, 
so $a^2 H \!=\! a_1^2 H_I$, where $a_1$ refers to the value of the 
scale factor at the end of inflation/beginning of radiation.
From~\cite{Glavan:2015cut} we have for the IR limit 
($k \!<\! \mu a H \! \ll \! aH$) of the modulus squared of the scalar 
mode function,
\begin{equation}
|\varphi_R(t,k)|^2 \approx \frac{4^{\nu_I-1}}{\pi} \Gamma^2(\nu_I)
	\bigl( \nu_I \!-\! \tfrac{1}{2} \bigr)^2 \frac{1}{a_1^3 H_I} 
	\Bigl( \frac{k}{a_1H_I} \Bigr)^{-2\nu_I}
	\Bigl[ 1 - \frac{1}{10} \Bigl( \frac{m}{H} \Bigr)^{\!2} \, \Bigr] \, ,
\end{equation}
where the expression is valid up to the first subleading order in small 
mass. The other combinations of the mode function appearing in the 
stochastic sources can be determined from taking derivatives of the 
expression above, and by using the equation of motion \eqref{modeEOM}. 
In particular,
\begin{equation}
|\dot{\varphi}_R(t,k)|^2
	= \frac{1}{2} \frac{d^2}{dt^2} |\varphi_R(t,k)|^2
		+ \frac{3H}{2} \frac{d}{dt} |\varphi_R(t,k)|^2 
		+ m^2 |\varphi_R(t,k)|^2 \, ,
\end{equation}
where the gradient terms were dropped. The stochastic sources 
during radiation period, to leading order in 
$0 \! \le\! |\xi|, (m/H)^2 \! \ll \! 1$, are, 
\begin{equation}
n_{\phi\phi}(N) \approx - \frac{H_I^2}{4\pi^2}  
	\biggl[ 1 - \frac{1}{10} \Bigl( \frac{m}{H} \Bigr)^{\!2} \,  \biggr] \, ,
\quad
n_{\phi\pi}(N) \approx  \frac{H_I^2}{10\pi^2}  
	\Bigl( \frac{m}{H} \Bigr)^{\!2}  \, ,
\quad
n_{\pi\pi}(N) \approx
	\mathcal{O}\Bigl( \frac{m}{H} \Bigr)^{\!\!4}  \, .
\end{equation}
The order $(m/H)^2$ to which the mode function was computed 
in~\cite{Glavan:2015cut} does not allow us to compute the leading 
order contribution to the last stochastic source above, but this is not 
important. Stochastic sources are relevant in inflation since they 
determine the amplitude of the correlators, but can actually safely 
be neglected in the subsequent periods of expansion. This argument 
can easily be made by examining the equations of motion,
\begin{eqnarray}
\frac{d}{d N} \Delta_{\phi\phi} - \Delta_{\phi\pi}  & = & n_{\phi\phi} \, ,
\label{radFULLeomPhiPhi}
\\
\frac{d}{d N} \Delta_{\phi\pi}
	+ \Delta_{\phi\pi} - 2 \Delta_{\pi\pi}
	+ 2 \Bigl(\frac{m}{H} \Bigr)^{\!2} \Delta_{\phi\phi} & = & n_{\phi\pi}  \, ,
\label{radFULLeomPhiPi}
\\
 \frac{d}{d N} \Delta_{\pi\pi} + 2 \Delta_{\pi\pi}
	+ \Bigl(\frac{m}{H} \Bigr)^{\!2} \Delta_{\phi\pi} & = & n_{\pi\pi}  \, ,
\label{radFULLeomPiPi}
\end{eqnarray}
and by comparing the relative size of various terms. The final values 
of correlators at the end of inflation~\eqref{minimal end of infl} 
or~\eqref{massless end of infl} serve as the initial conditions for the 
radiation period, and we note that both in minimally coupled, or 
massless limit they contain enhancement factors of~$N_I$ 
or~$e^{8|\xi|N_I}\!/|\xi|$, respectively. These derive partly from 
expanding the IR phase space of modes included in long wavelength 
correlators. On the other hand, none of the sources on the right hand 
side of~\eqref{radFULLeomPhiPhi}--\eqref{radFULLeomPiPi} has this 
enhancement factor, making them negligible in comparison.
Hence, to a very good approximation, the equations for IR 
correlators during (late) radiation era are the same as the classical ones,
\begin{eqnarray}
\frac{d}{d N} \Delta_{\phi\phi} - \Delta_{\phi\pi}  & \approx & 0 \, ,
\label{radEOM1}
\\
\frac{d}{d N} \Delta_{\phi\pi}
	+ \Delta_{\phi\pi} - 2 \Delta_{\pi\pi}
	+ 2 \Bigl(\frac{m}{H} \Bigr)^{\!2} \Delta_{\phi\phi} & \approx & 0  \, ,
\label{radEOM2}
\\
 \frac{d}{d N} \Delta_{\pi\pi} + 2 \Delta_{\pi\pi}
	+ \Bigl(\frac{m}{H} \Bigr)^{\!2} \Delta_{\phi\pi} & \approx & 0  \, .
\label{radEOM3}
\end{eqnarray}
These equations we may solve perturbatively in small mass (and 
small nonminimal coupling contained in the initial conditions).
To do this properly we need to take into account that the initial 
conditions of the radiation period satisfy the hierarchy inherited 
from inflation,
\begin{equation}
\Delta_{\phi\phi}(0) \gg \Delta_{\phi\pi}(0) \gg \Delta_{\pi\pi}(0) \, ,
\end{equation}
which is dictated by the same perturbative parameters. Therefore 
the equations that correctly capture the leading order correlators 
(denoted by superscript $(0)$) are
\begin{eqnarray}
\frac{d}{dN}\Delta_{\phi\phi}^{(0)} & = & 0 \, ,
\\
\frac{d}{d N} \Delta_{\phi\pi}^{(0)} + \Delta_{\phi\pi}^{(0)} 
	+ 2 \Bigl(\frac{m}{H} \Bigr)^{\!2} \Delta_{\phi\phi}^{(0)} & =& 0  \, ,
\\
 \frac{d}{d N} \Delta_{\pi\pi}^{(0)} + 2 \Delta_{\pi\pi}^{(0)}
	+ \Bigl(\frac{m}{H} \Bigr)^{\!2} \Delta_{\phi\pi}^{(0)} & = & 0  \, ,
\end{eqnarray}
which we can readily solve,
\begin{eqnarray}
\Delta_{\phi\phi}^{(0)}(N) & = & \Delta_{\phi\phi}(0) \, ,
\label{Delta phiphi0:R}
\\
\Delta_{\phi\pi}^{(0)}(N) & = & \Delta_{\phi\pi}(0) e^{-N}
	- \frac{2}{5} \Delta_{\phi\phi}(0) \Bigl( \frac{m}{H} \Bigr)^{\!2}
		\Bigl[ 1 \!-\! e^{-5N} \Bigr] \, ,
\label{Delta pipi0:R}\\
\Delta_{\pi\pi}^{(0)}(N) & = & \Delta_{\pi\pi}(0) e^{-2N}
	+ \frac{1}{25} \Delta_{\phi\phi}(0) \Bigl( \frac{m}{H} \Bigr)^{\!4}
	\Bigl[ 1 \!-\! e^{-5N} \Bigr]^2
	- \frac{1}{5}  \Delta_{\phi\pi}(0) \Bigl( \frac{m}{H} \Bigr)^{\!2}
	\Bigl[ 1 \!-\! e^{-5N} \Bigr] e^{-N} \, . \qquad
\label{Delta pipi0:R}
\end{eqnarray}
This result encompasses both minimally coupled and massless limits.

At the end of radiation period (when $N_R\sim 50$)
 we can further simplify~(\ref{Delta phiphi0:R}--\ref{Delta pipi0:R})
by dropping the terms that have exponentially decayed,
\begin{eqnarray}
\Delta_{\phi\phi}^{(0)}(N_R) & = & \Delta_{\phi\phi}(0) \, ,
\label{RADfinalPhiPhi}
\\
\Delta_{\phi\pi}^{(0)}(N_R) & = & \Delta_{\phi\pi}(0) e^{-N_R}
	- \frac{2}{5} \Delta_{\phi\phi}(0) 
		\Bigl( \frac{m}{H_{\rm eq}} \Bigr)^{\!2} \, ,
\label{RADfinalPhiPi}
\\
\Delta_{\pi\pi}^{(0)}(N_R) & = & \Delta_{\pi\pi}(0) e^{-2N_R}
	+ \frac{1}{25} \Delta_{\phi\phi}(0) 
		\Bigl( \frac{m}{H_{\rm eq}} \Bigr)^{\!4} \, ,
\label{RADfinalPiPi}
\end{eqnarray}
where $N_R\!\approx\!50$ is the total number of e-foldings during 
radiation,~\footnote{This estimate assumes inflation at the 
scale $H_I\sim 10^{13}\!~\!{\rm GeV}$. Since in general 
$N_R\gg 1$, our estimates apply also to lower scale inflationary models.}
and $H_{\rm eq}$ is the Hubble rate at the end of radiation and beginning 
of matter period (radiation-matter equality). The energy density and 
pressure to leading order in radiation period are
\begin{eqnarray}
\rho_Q \!&=&\! \frac{3 H_I^4}{32\pi^2} e^{8|\xi|N_I} 
	\biggl[ - \Bigl( \frac{H}{H_I} \Bigr)^2 
		+ \frac{1}{6|\xi|} \Bigl( \frac{m}{H_I} \Bigr)^2 \biggr]\, , 
\label{rho Q: matter era}\\
p_Q \!&=&\! \frac{3 H_I^4}{32\pi^2} e^{8|\xi|N_I} 
	\biggl[ - \frac{1}{3} \Bigl( \frac{H}{H_I} \Bigr)^2  
		- \frac{1}{6|\xi|} \Bigl( \frac{m}{H_I} \Bigr)^2 \biggr] \, ,
\label{p Q: matter era}
\end{eqnarray}
which is exactly what was obtained in~\cite{Glavan:2015cut}. There 
are two competing contributions 
in~(\ref{rho Q: matter era}--\ref{p Q: matter era}): a radiation-like 
fluid with negative energy and pressure, and a cosmological constant 
type contribution.


\subsection{Matter period}

For the matter period the correlator values at the end of 
radiation~\eqref{RADfinalPhiPhi}-\eqref{RADfinalPiPi} serve as initial
conditions.
Applying the same reasoning as in the radiation period case,
the leading order equations of motion for the IR correlators in 
matter period  are
\begin{eqnarray}
\frac{d}{dN} \Delta_{\phi\phi}^{(0)} & \approx & 0 \, ,
\label{matter correlator I}\\
\frac{d}{dN} \Delta_{\phi\pi}^{(0)} + \frac{3}{2} \Delta_{\phi\pi}^{(0)}
	+ 2 \biggl[ \Bigl( \frac{m}{H} \Bigr)^{\!2} 
	+ 3\xi \biggr] \Delta_{\phi\phi}^{(0)} &  \approx & 0 \, ,
\label{matter correlator II}\\
\frac{d}{dN} \Delta_{\pi\pi}^{(0)} + 3 \Delta_{\pi\pi}^{(0)}
	+  \biggl[ \Bigl( \frac{m}{H} \Bigr)^{\!2}  
	+ 3\xi \biggr] \Delta_{\phi\pi}^{(0)} &  \approx & 0 \, ,
\label{matter correlator III}
\end{eqnarray}
whose solutions are
\begin{eqnarray}
\Delta_{\phi\phi}^{(0)}(N) & = & \Delta_{\phi\phi}(0) \, ,
\\
\Delta_{\phi\pi}^{(0)}(N) & = & \Delta_{\phi\pi}(0) e^{-3N/2}
	- 4\xi \Delta_{\phi\phi}(0) \Bigl[ 1 \!-\! e^{-3N/2} \Bigr]
	- \frac{4}{9} \Delta_{\phi\phi}(0) \Bigl( \frac{m}{H} \Bigr)^{\!2}
		\Bigl[ 1 \!-\! e^{-9N/2} \Bigr] \, ,
\\
\Delta_{\pi\pi}^{(0)}(N) & = & \Delta_{\pi\pi}(0) e^{-3N}
	- \frac{2}{9} \Delta_{\phi\pi}(0) \Bigl( \frac{m}{H} \Bigr)^{\!2}
		\Bigl[ 1 \!-\! e^{-9N/2} \Bigr] e^{-3N/2}
	- 2\xi \Delta_{\phi\pi}(0) \Bigl[ 1 \!-\! e^{-3N/2} \Bigr] e^{-3N/2}
\nonumber \\
& & + \frac{4}{81} \Delta_{\phi\phi}(0) \biggl[ 
		\Bigl( \frac{m}{H} \Bigr)^{\!2}
		\Bigl[ 1 \!-\! e^{-9N/2} \Bigr] 
		+ 9\xi \Bigl[ 1 \!-\! e^{-3N/2} \Bigr] \biggr]^{2} \, .
\end{eqnarray}
Matter period lasts for about $N_M\!\approx\!8$ e-foldings before 
the onset of DE domination, so we can further simplify the results above,
\begin{eqnarray}
\Delta_{\phi\phi}(N_M) & = & \Delta_{\phi\phi}(0) \, ,
\\
\Delta_{\phi\pi}(N_M) & = & 4 \biggl[ |\xi| 
	-  \frac{1}{9}\Bigl( \frac{m}{H} \Bigr)^{\!\!2} \, \biggr] 
		\Delta_{\phi\phi}(0) \, ,
\\
\Delta_{\pi\pi}(N_M) & = & 4 \biggl[  |\xi| 
	- \frac{1}{9} \Bigl( \frac{m}{H} \Bigr)^{\!\!2} \, \biggr]^2
		 \Delta_{\phi\phi}(0) \, ,
\end{eqnarray}
where $H_{\rm eq}>H\gg H_{DE}$
and  $H_{DE}$ is the Hubble rate at the onset of DE domination,
{\it i.e.} when $\rho_c=\rho_Q$.
We see that the evolution preserves the hierarchy between the 
correlators induced by inflation, as long as the mass is smaller than 
the evolving Hubble rate. These we take as initial values for numerical 
evolution of the self-consistent quantum-corrected Friedmann equations.

The energy density and pressure in matter era has the same form 
as during radiation,
\begin{equation}
\rho_Q = \frac{3 H_I^4}{32\pi^2} e^{8|\xi|N_I} 
	\biggl[ - \Bigl( \frac{H}{H_I} \Bigr)^2 
	+ \frac{1}{6|\xi|} \Bigl( \frac{m}{H_I} \Bigr)^2 \biggr]\, , 
\qquad
p_Q = \frac{3 H_I^2}{32\pi^2} e^{8|\xi|N_I} 
	\biggl[  - \frac{1}{6|\xi|} \Bigl( \frac{m}{H_I} \Bigr)^2 \biggr] \, ,
\label{mat rho p}
\end{equation}
confirming the results of~\cite{Glavan:2015cut}. Note that this expression 
is derived as a leading order contribution in the limit $0<-\xi\ll1$ and 
$(m/H)^2\ll1$. It does not quantitatively capture the limit $\xi\!=\!0$, 
in which case the enhancement factor $e^{8|\xi| N_I}$ takes a different 
form. Also, the case of a heavy scalar $(m/H)^2\gtrsim 1$ is not captured 
by these expressions, where the energy density and pressure of the scalar 
are of the nonrelativistic matter form at leading order.

There are two competing 
contributions in~(\ref{mat rho p}), one behaving like nonrelativistic 
matter (with negative energy), and the other one like a cosmological 
constant (CC). If the parameters of the model are such that the CC-like 
term starts dominating over the matter term before late times in
matter era it should also eventually dominate over the background fluid 
and accelerate the expansion. This regime is beyond the (perturbative) 
regime discussed in the present section, where the backreaction on the 
expansion rate was neglected. It requires solving the set of semiclassical 
Friedman equations~(\ref{Friedman eqs}) (with 
$\rho_c\rightarrow \rho_c+\rho_Q$ and $p_c\rightarrow p_c+p_Q$)
together with the matter era correlator equations self-consistently, 
as described at the end of section~\ref{sec: Stochastic formalism}.
The expansion dynamics during that regime are determined in the 
next section.


\section{Self-consistent solution and results}

Here we present solutions of the semiclassical Friedman equations 
with the initial conditions set during matter era at time $t=t_*$ when 
still $H(t_*)=H_*\gg H_{ DE}$. The quantum backreaction
energy density and pressure~(\ref{backreactionRHO}--\ref{backreactionP}) 
represent sources for the Friedman equations~(\ref{Friedman eqs})
on top of the classical matter fluid. The equations of motion for the 
correlators~(\ref{eomPhiPhi}--\ref{eomPiPi}) with the sources neglected
close the full set of equations.

\begin{figure}[h!]
\vskip -1cm
\begin{minipage}[b]{8cm}
$[m/H(z_{in})]^2 = 10^{-4}, \ \xi= 0$
\vskip+0.3cm
    \includegraphics[width=7.5cm] {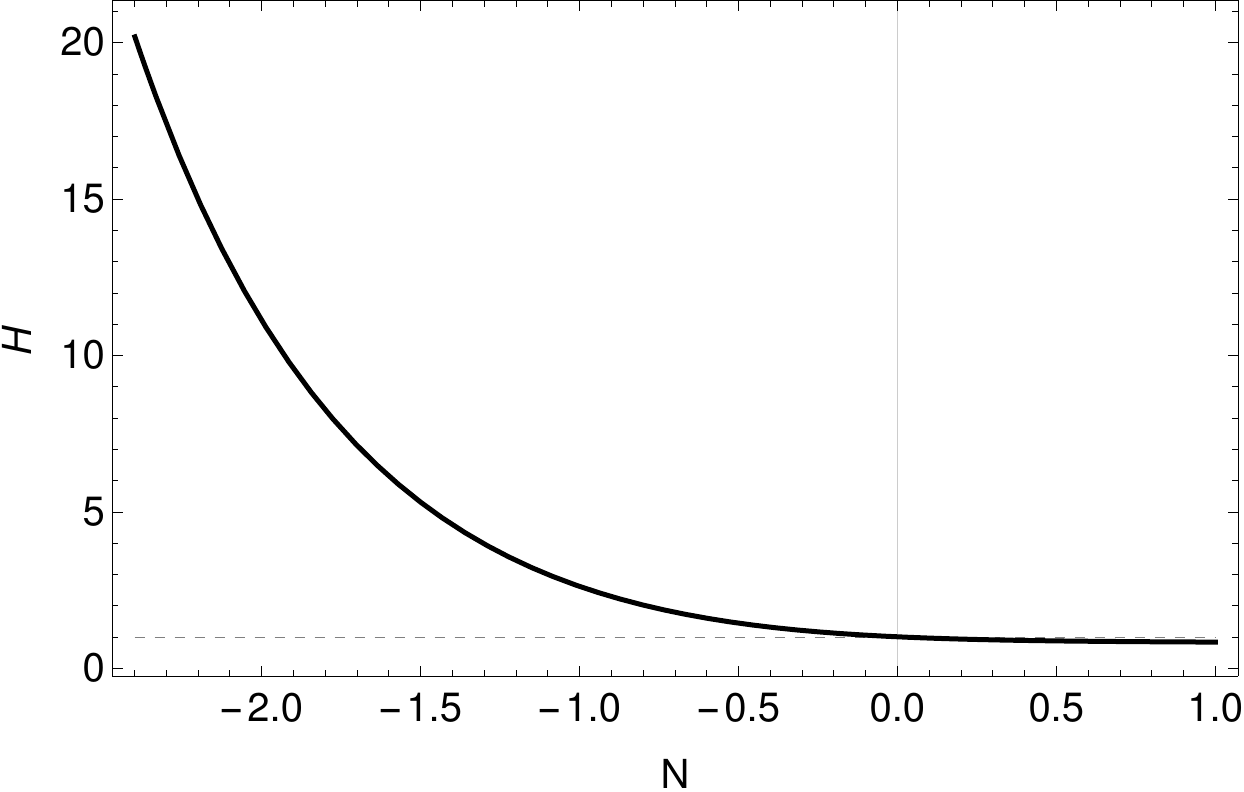}
\end{minipage}
\hfill
\begin{minipage}[b]{8cm}
$[m/H(z_{in})]^2 = 10^{-3}, \ \xi= 0$
\vskip+0.3cm
    \includegraphics[width=7.5cm] {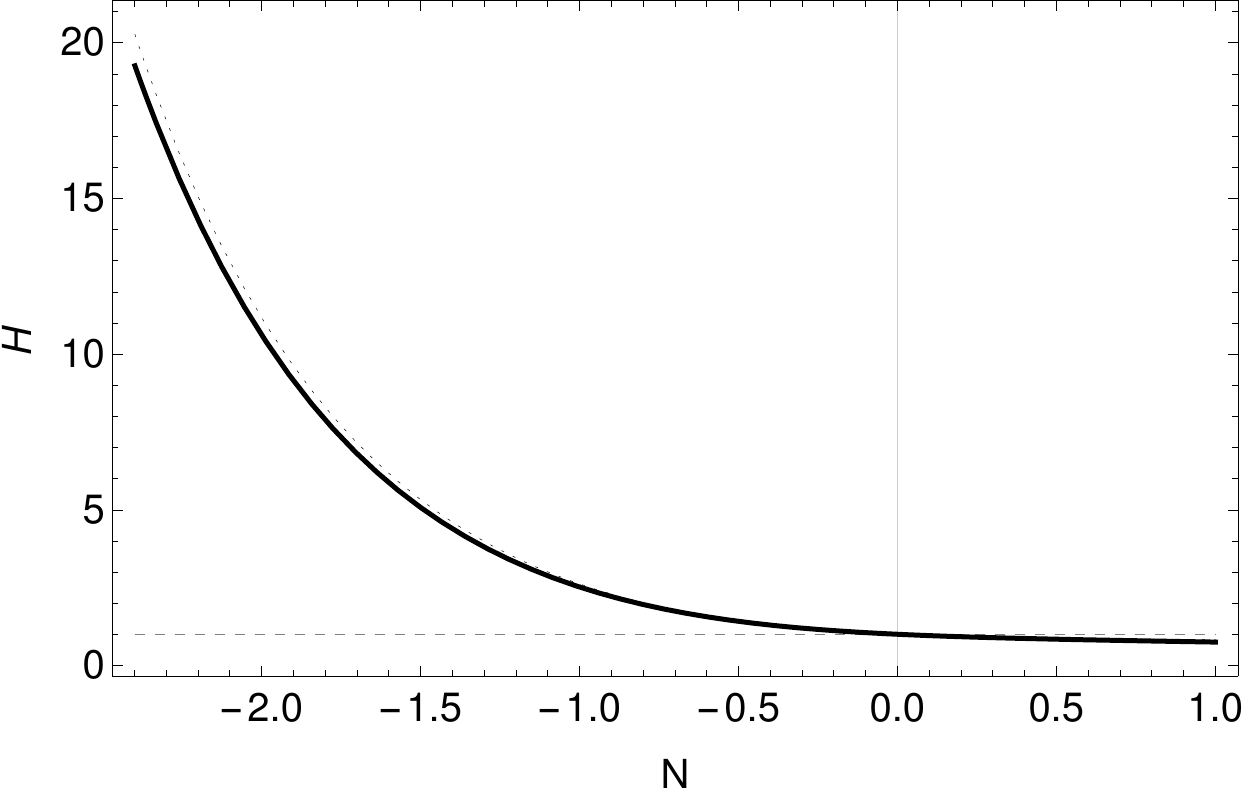}
\end{minipage}
\vskip+0.3cm
\begin{minipage}[b]{8cm}
    \includegraphics[width=7.5cm] {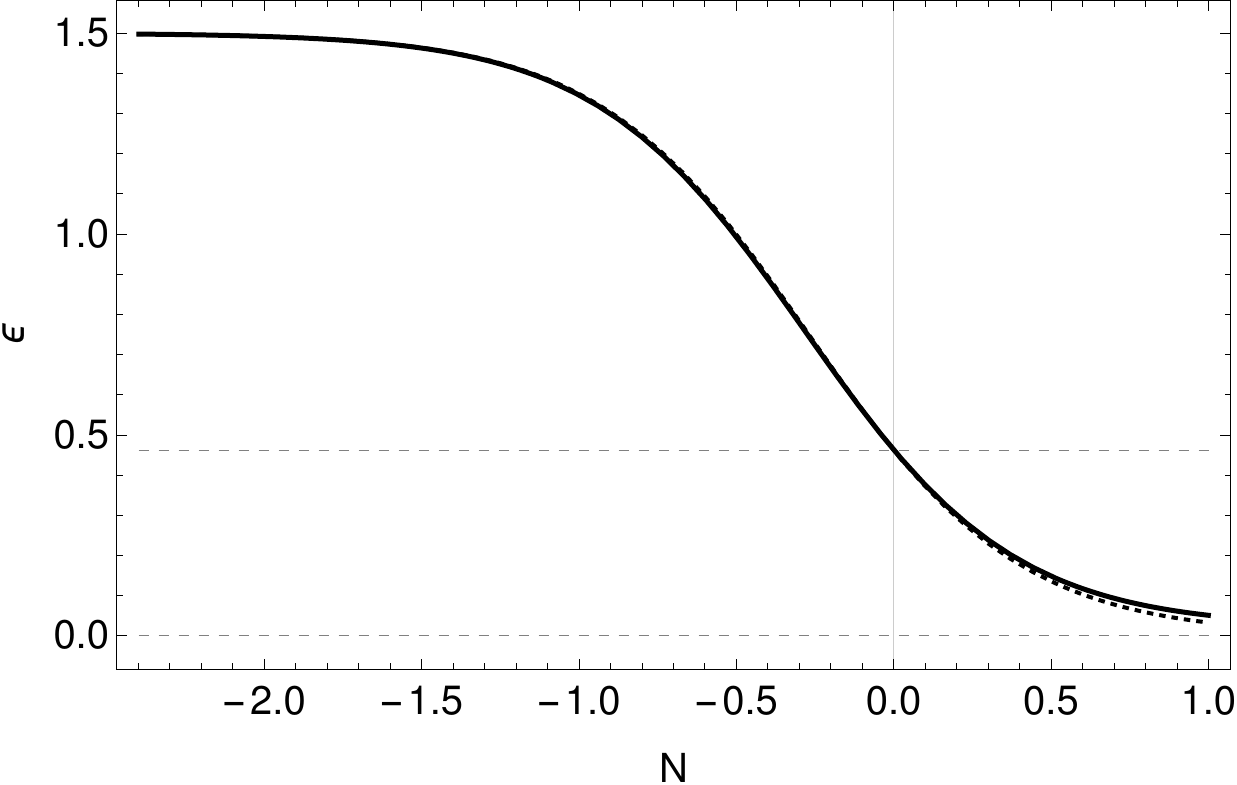}
\end{minipage}
\hfill
\begin{minipage}[b]{8cm}
    \includegraphics[width=7.5cm] {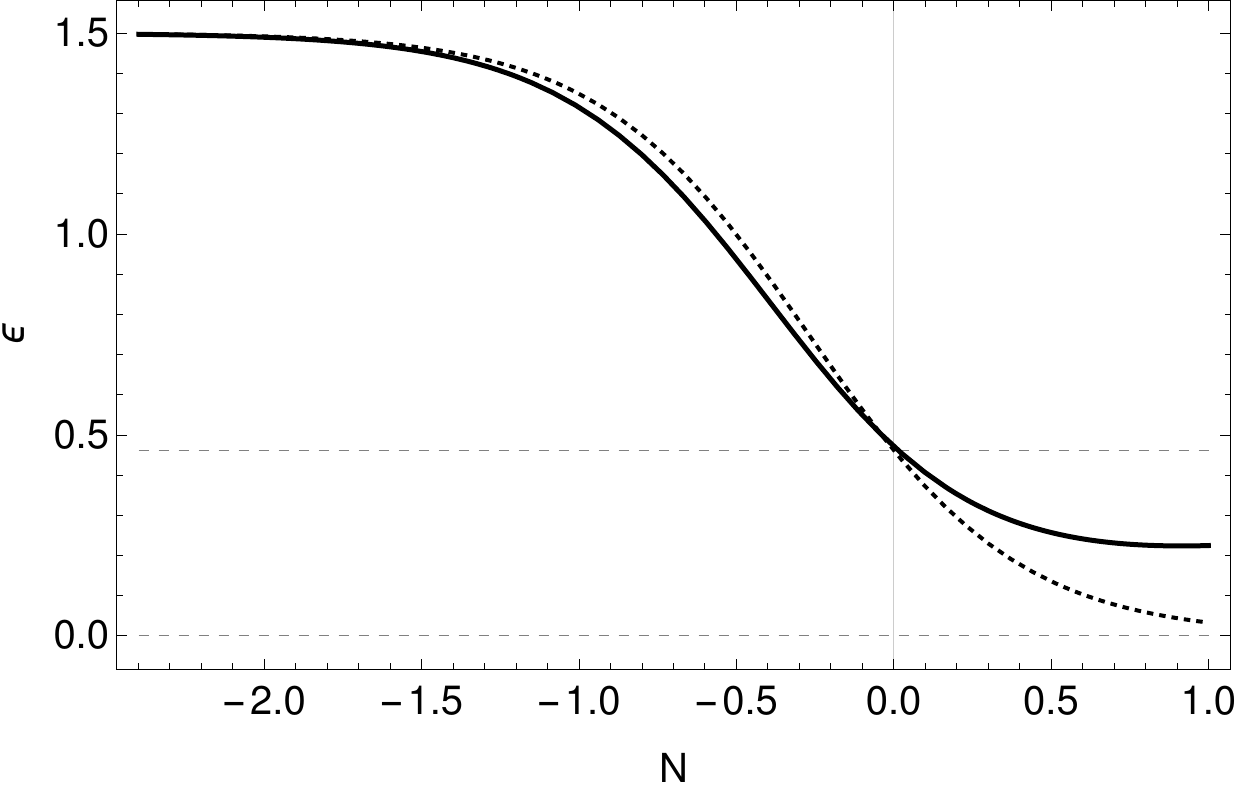}
\end{minipage}
\vskip+0.3cm
\begin{minipage}[b]{8cm}
    \includegraphics[width=7.5cm] {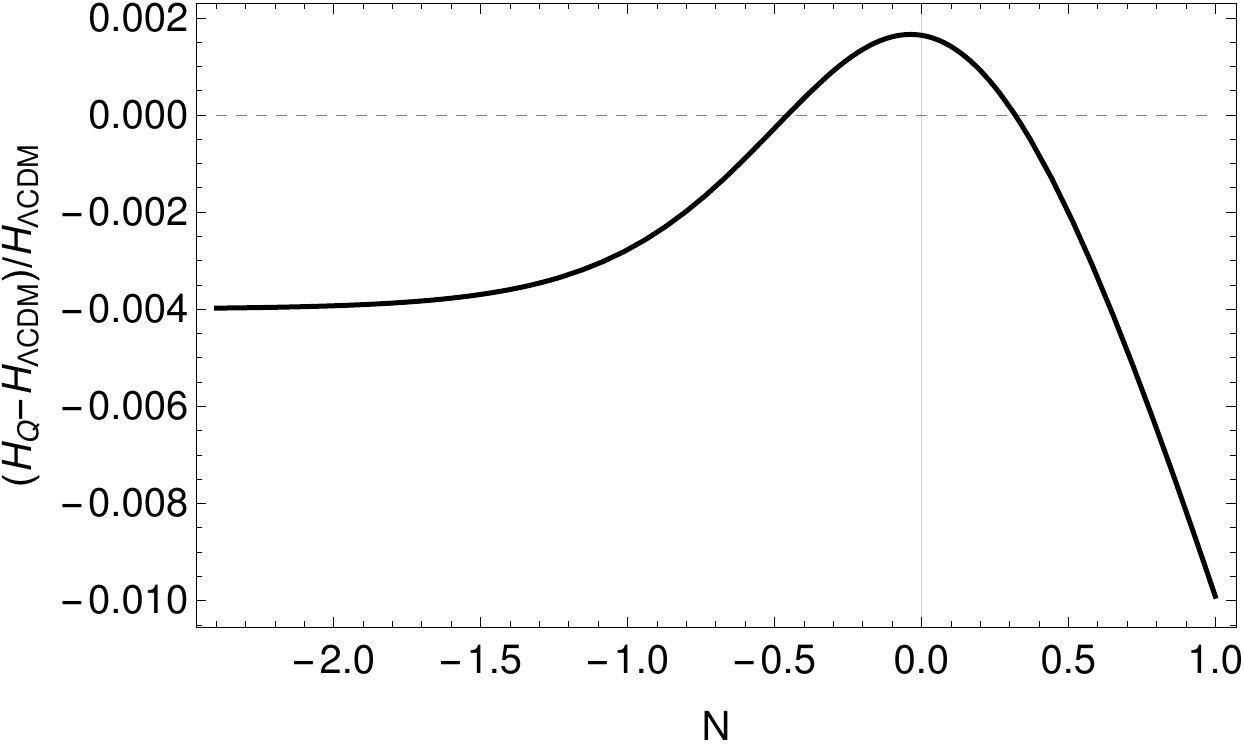}
\end{minipage}
\hfill
\begin{minipage}[b]{8cm}
     \includegraphics[width=7.5cm] {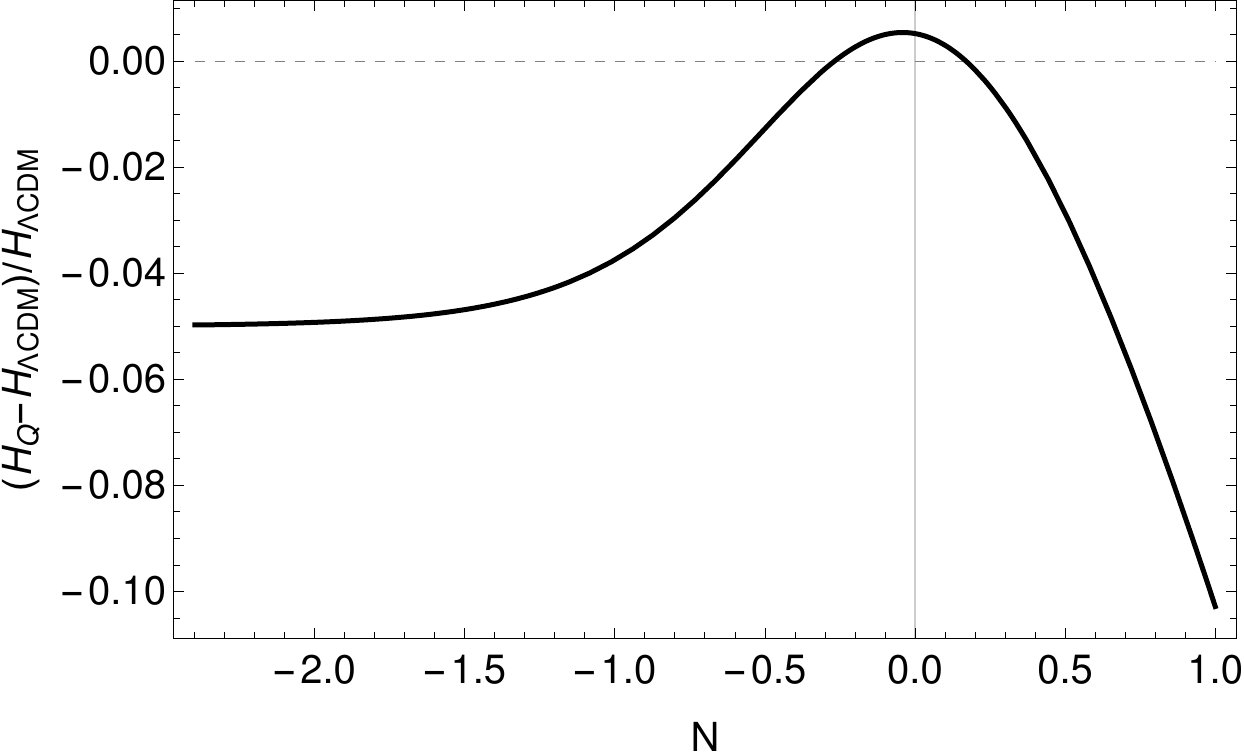}
\end{minipage}
\vskip+0.3cm
\begin{minipage}[b]{8cm}
    \includegraphics[width=7.5cm] {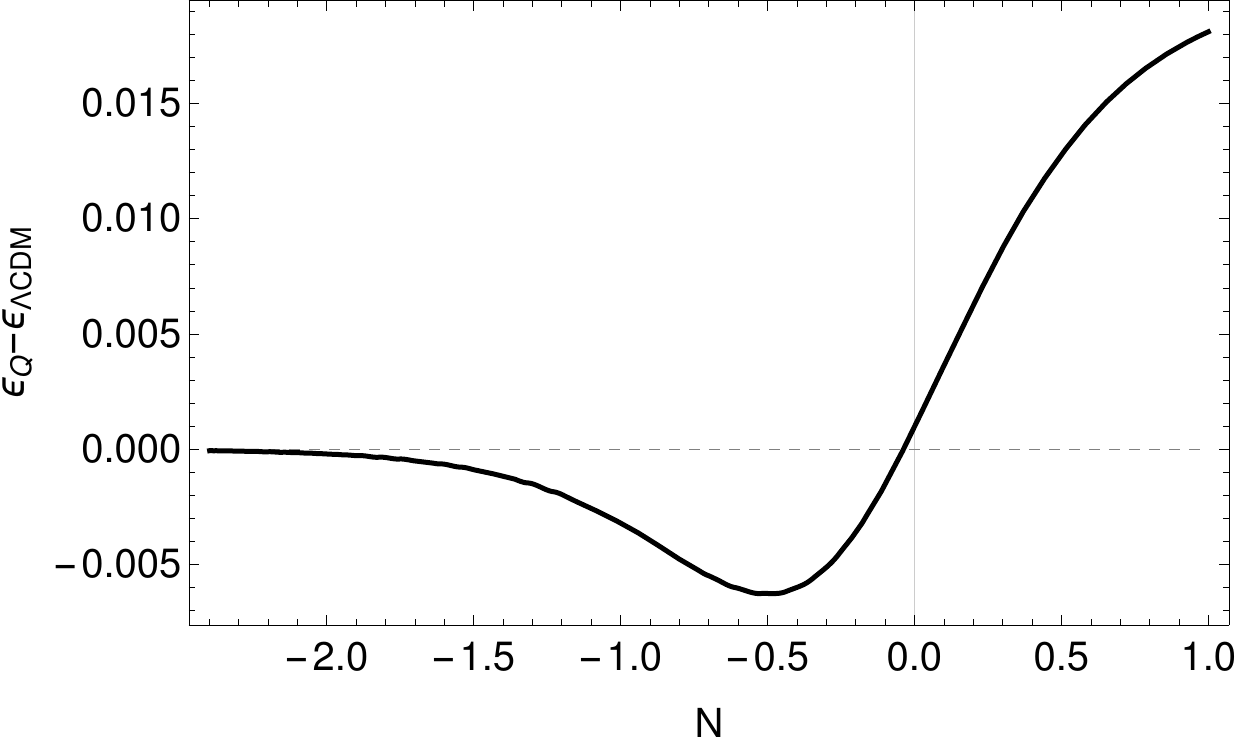}
\end{minipage}
\hfill
\begin{minipage}[b]{8cm}
     \includegraphics[width=7.5cm] {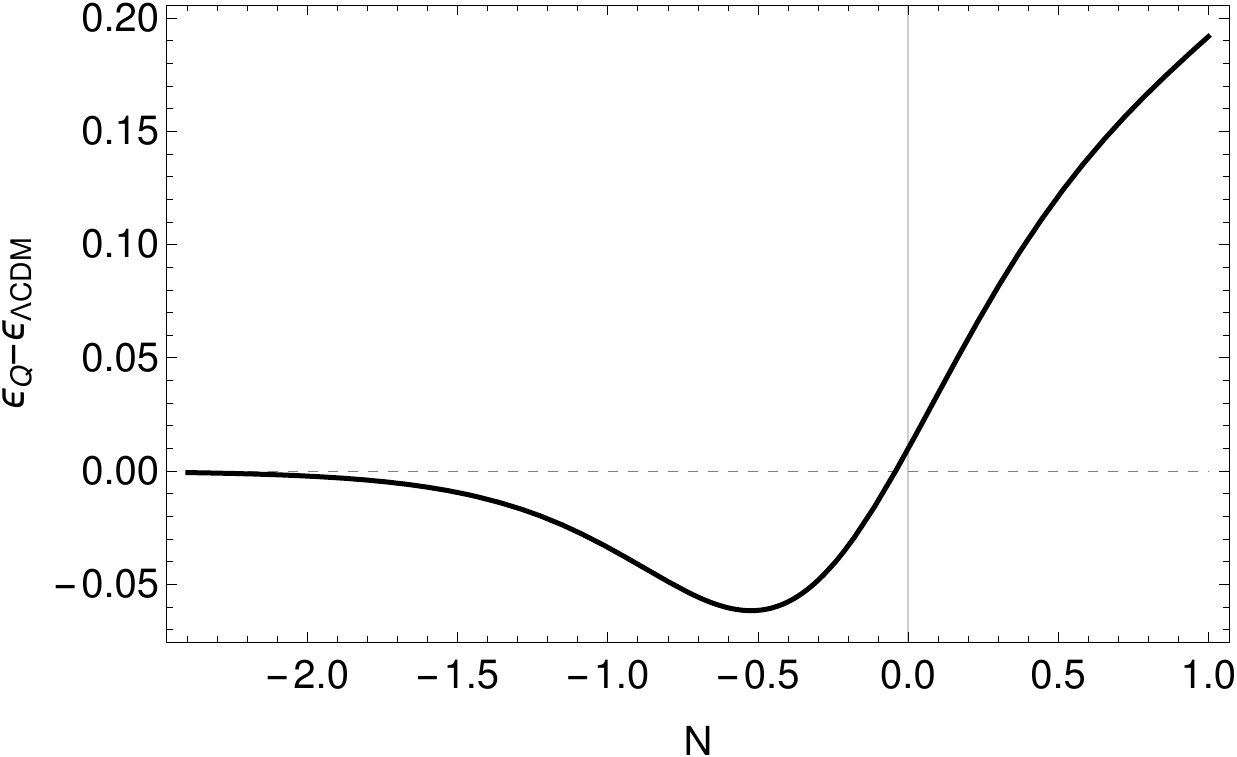}
\end{minipage}
\vskip-0.4cm
\caption{Quantum backreaction model versus the $\Lambda$CDM 
model. Two columns represent two different choices of parameters 
$\xi$ and $m$.
{\it Left column:} $[m/H(z_{in})]^2 \!=\! 10^{-4}$, $\xi \!=\! 0$.
{\it Right column:} $[m/H(z_{in})]^2 \!=\! 10^{-3}$, $\xi \!=\! 0$. 
The first row shows plots of the Hubble rate, and the second row 
of the $\epsilon$ parameter in the backreaction model (solid line), 
and in the $\Lambda$CDM (dotted line). The third and fourth rows 
illustrate the difference between the Hubble rate and $\epsilon$ 
in the backreaction model and the $\Lambda$CDM model, respectively.}
\label{results}
\end{figure}

\begin{figure}[h]
\vskip -1cm
\begin{minipage}[b]{8cm}
$[m/H(z_{in})]^2 = 10^{-4}, \ \xi= - 10^{-3}$
\vskip+0.3cm
    \includegraphics[width=7.5cm] {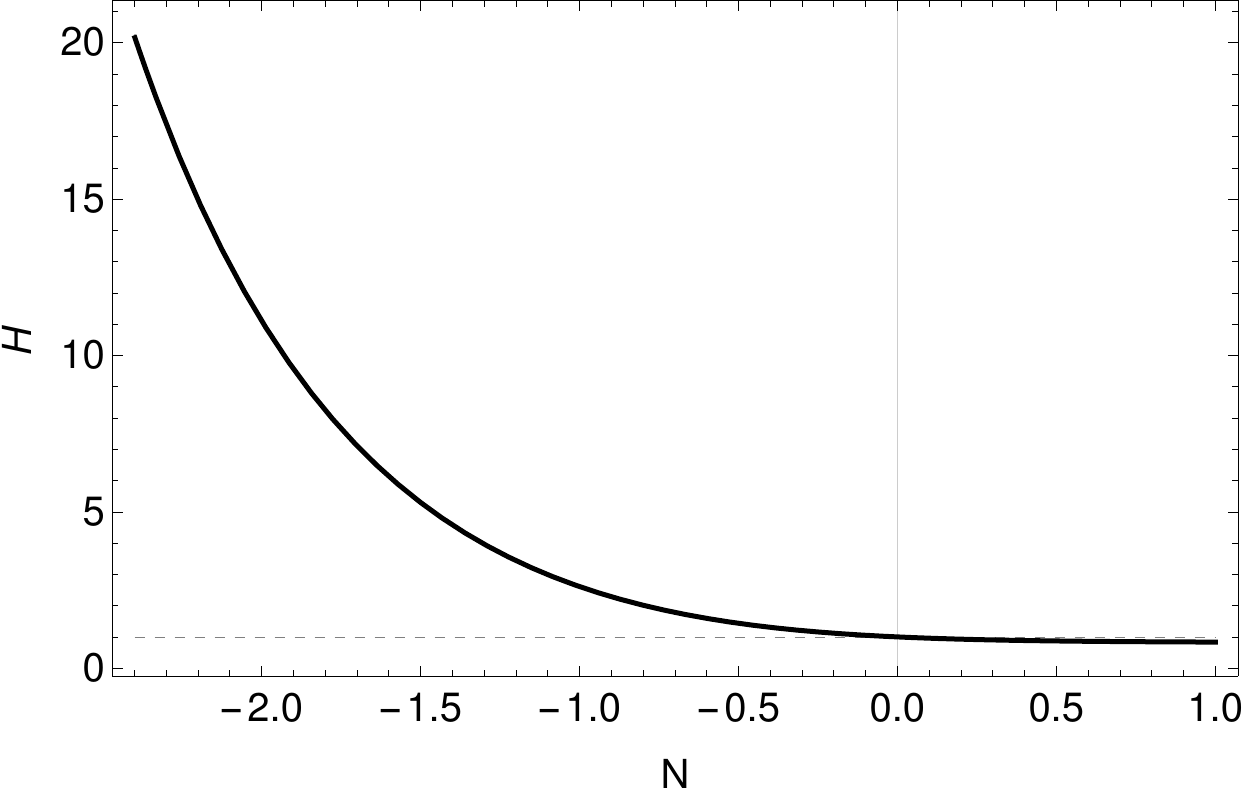}
\end{minipage}
\hfill
\begin{minipage}[b]{8cm}
$[m/H(z_{in})]^2 = 10^{-3}, \ \xi= -10^{-3}$
\vskip+0.3cm
    \includegraphics[width=7.5cm] {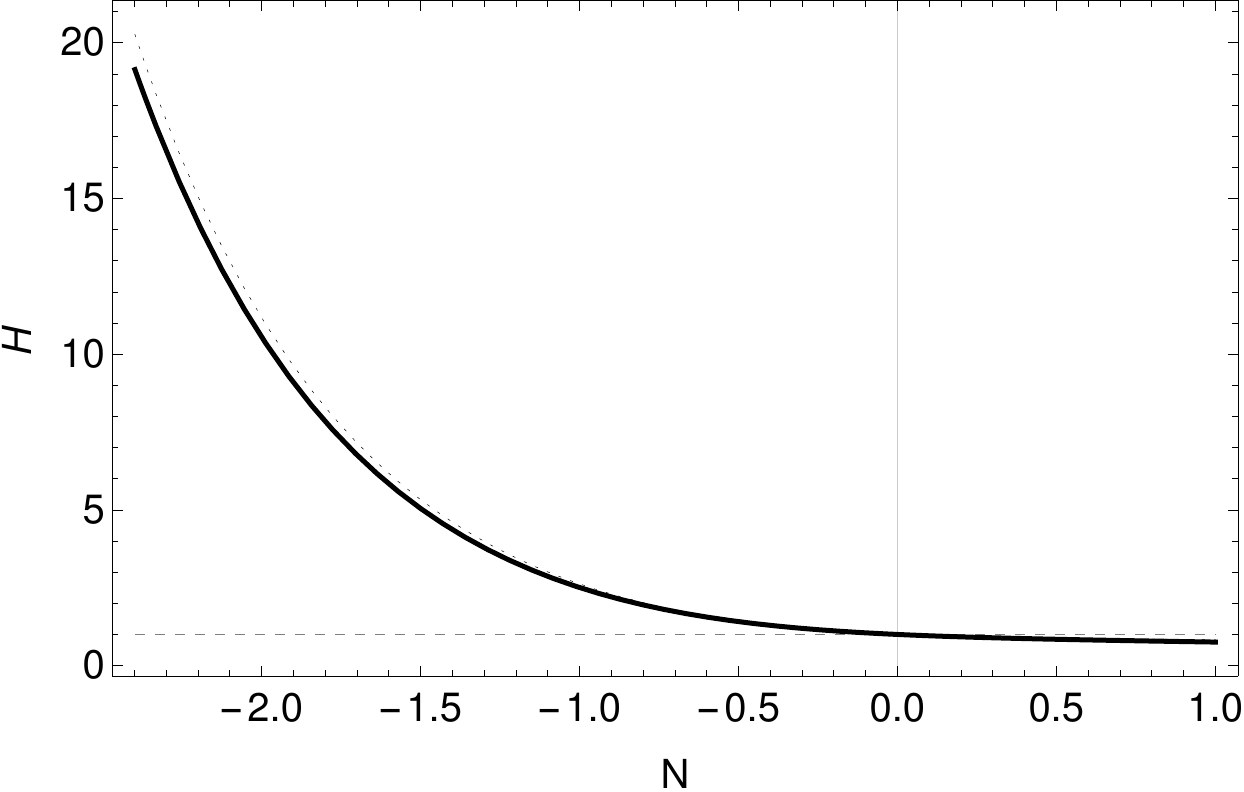}
\end{minipage}
\vskip+0.3cm
\begin{minipage}[b]{8cm}
    \includegraphics[width=7.5cm] {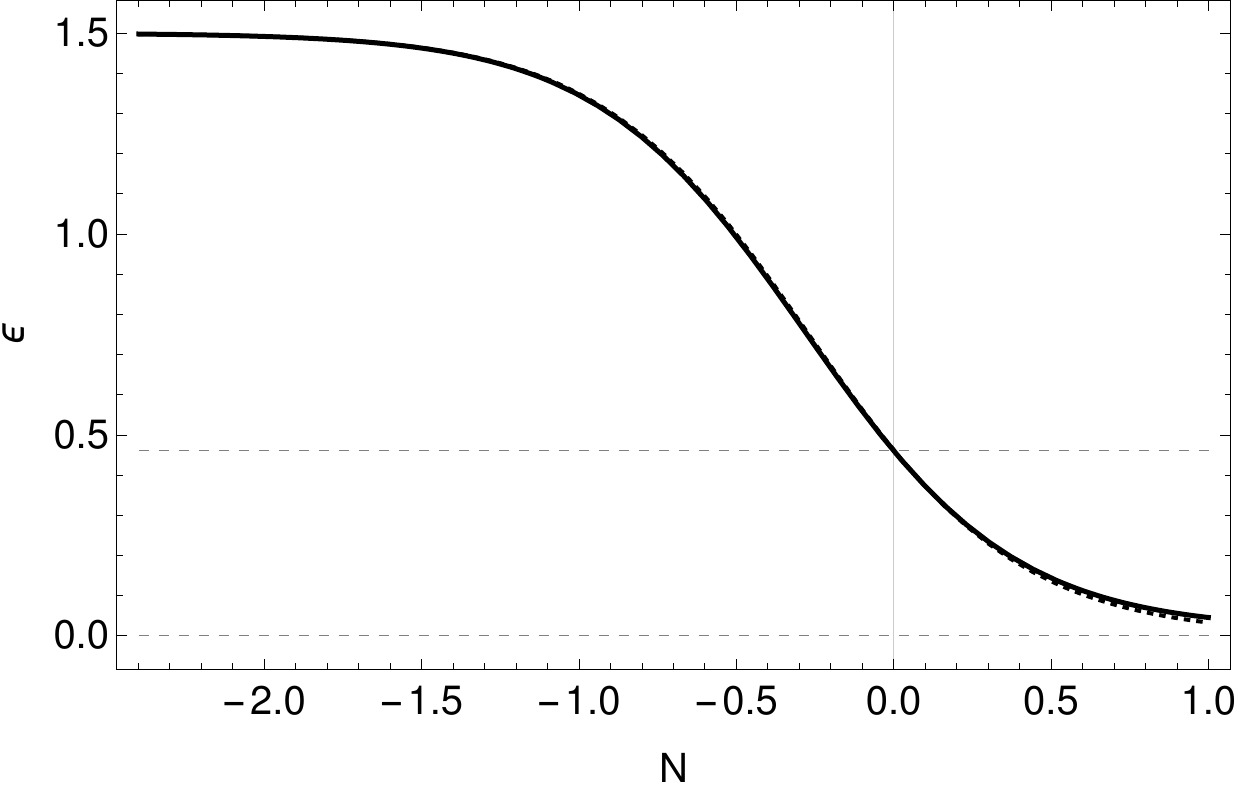}
\end{minipage}
\hfill
\begin{minipage}[b]{8cm}
    \includegraphics[width=7.5cm] {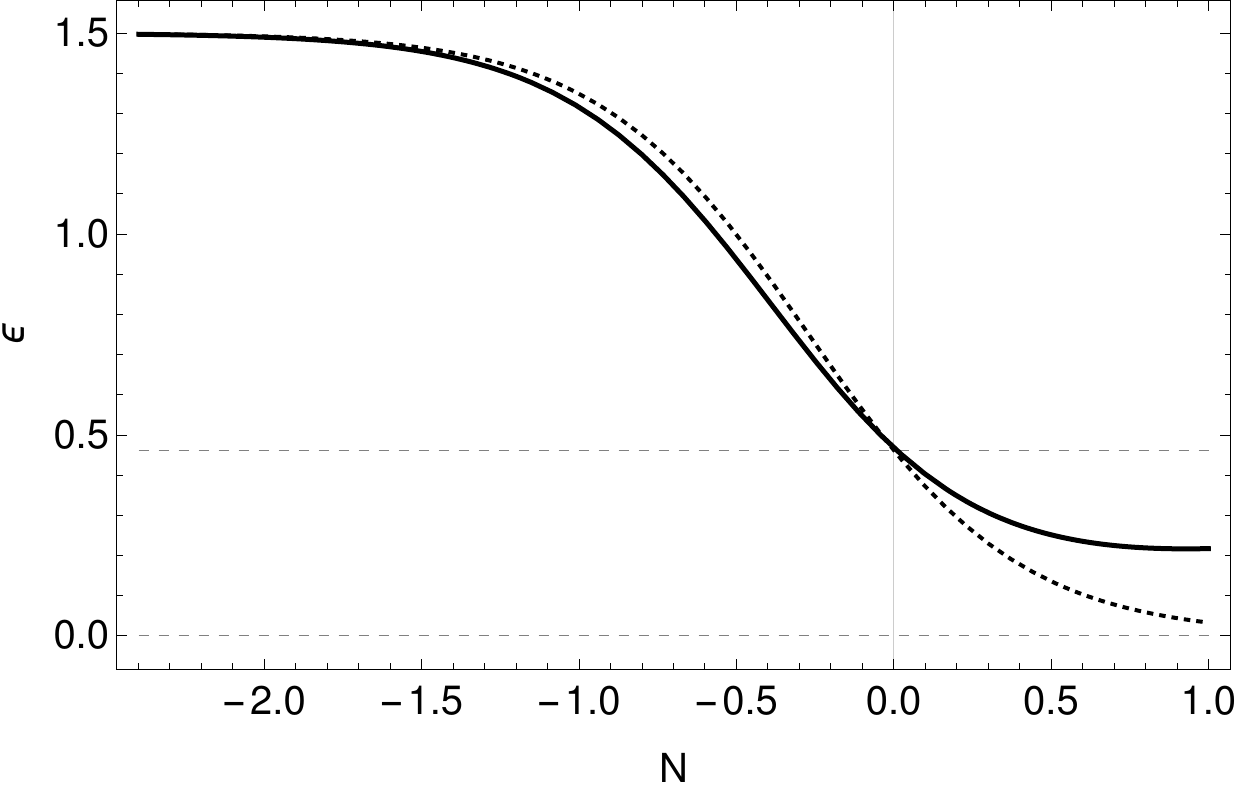}
\end{minipage}
\vskip+0.3cm
\begin{minipage}[b]{8cm}
    \includegraphics[width=7.5cm] {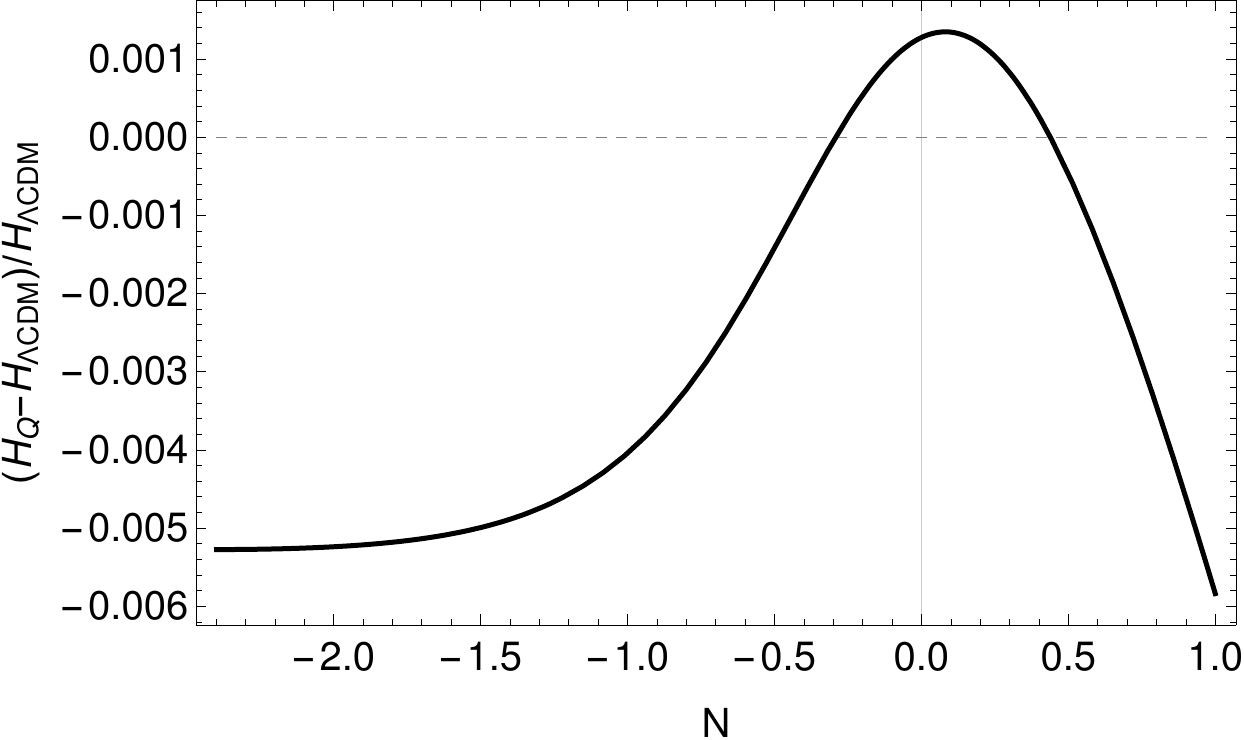}
\end{minipage}
\hfill
\begin{minipage}[b]{8cm}
    \includegraphics[width=7.5cm] {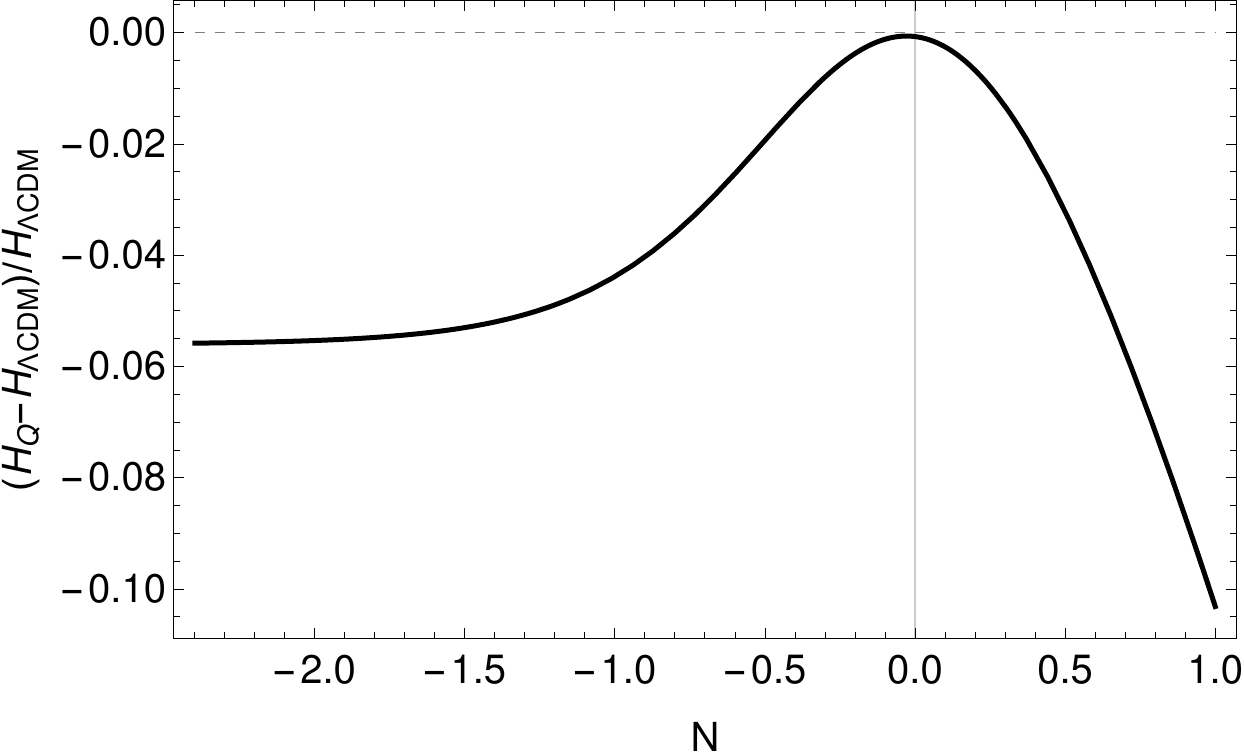}
\end{minipage}
\vskip+0.3cm
\begin{minipage}[b]{8cm}
    \includegraphics[width=7.5cm] {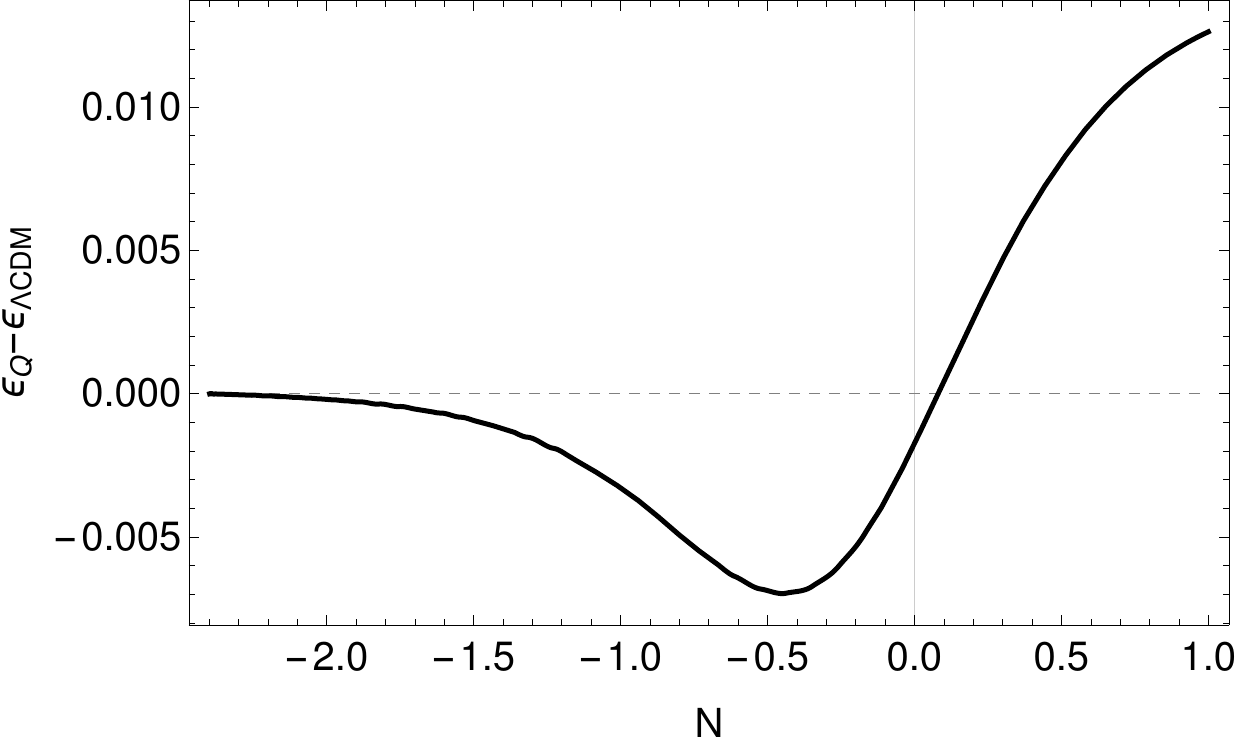}
\end{minipage}
\hfill
\begin{minipage}[b]{8cm}
    \includegraphics[width=7.5cm] {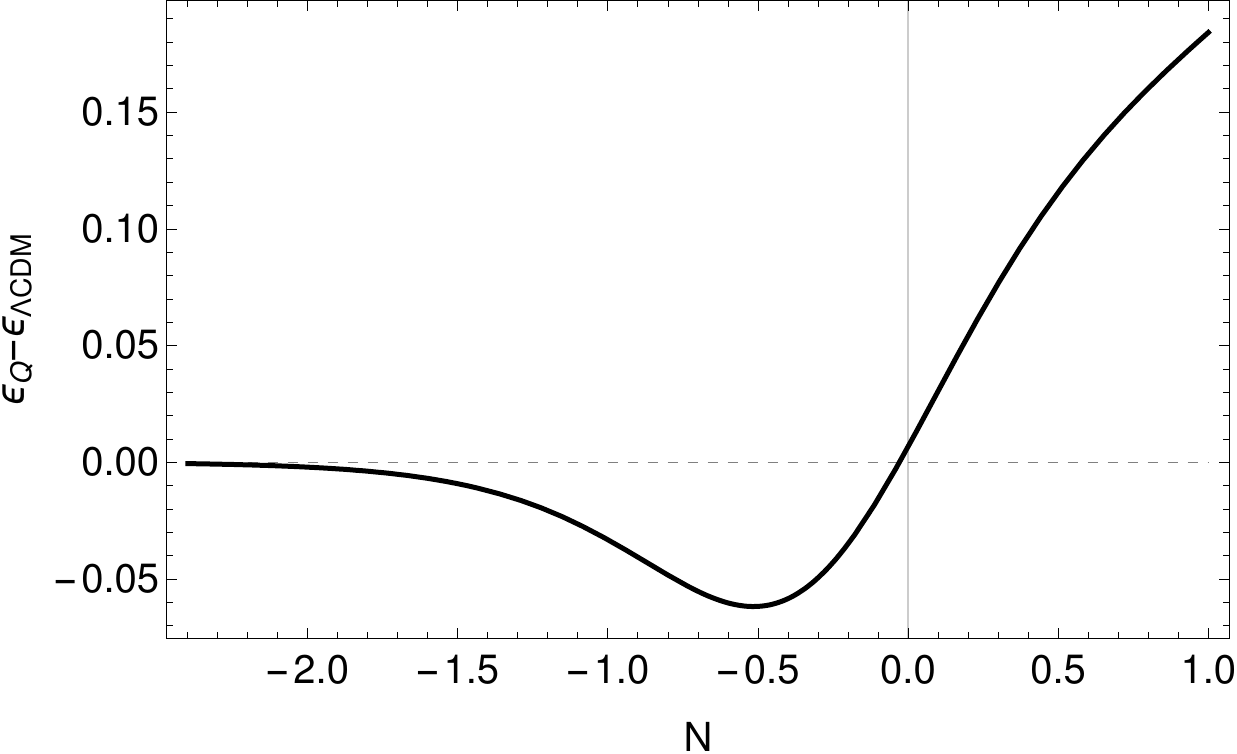}
\end{minipage}
\vskip-0.4cm
\caption{Quantum backreaction model versus the $\Lambda$CDM 
model. Two columns represent two different choices of parameters 
$\xi$ and $m$. {\it Left column:} $[m/H(z_{in})]^2 \!=\! 10^{-4}$, 
$\xi \!=\! -10^{-3}$. {\it Right column:} $[m/H(z_{in})]^2 \!=\! 10^{-3}$, 
$\xi \!=\! -10^{-3}$. The first row shows plots of the Hubble rate, and 
the second row of the $\epsilon$ parameter in the backreaction 
model (solid line), and in the $\Lambda$CDM (dotted line). 
The third and fourth rows illustrate the difference between 
the Hubble rate and $\epsilon$ in the backreaction model and 
the $\Lambda$CDM model, respectively.
 }
\label{results2}
\end{figure}
%

More specifically, initial conditions are set at redshift 
$z \!=\! z_{in} \!=\! 10$ (corresponding to about 2 e-foldings 
in the past from today, $N(z) \!=\!-\ln(1\!+\!z)$) in such a way 
that the energy density and pressure of the would-be 
cosmological constant is substituted for the quantum 
backreaction. The equations are then solved numerically.
The solutions for different choices of parameters are given 
in Figs.~\ref{results} and~\ref{results2} which show how 
geometric quantities such as the Hubble rate $H$ and 
$\epsilon \!=\! (3/2)(1\!+\!w)$ (vertical axis) depend on the 
number of e-foldings $N=-\ln({1\!+\!z})$ (horizontal axis).

In Fig.~\ref{results} we see that, when $\xi\!=\!0$ and 
$m \!=\! 0.20H_0$ [$m\!=\!0.01H(z\!=\!10)$], the relative difference 
between the Hubble expansion rate and $\epsilon$ in our model
and $\Lambda$CDM is at a sub-percent level, which is not observable 
by current observations and at the verge of being observable 
by the near future observatories.  However the differences become 
more pronounced when the scalar mass approaches the Hubble rate 
today. Indeed, when $m\simeq 0.61H_0$ [$m^2=10^{-3}H^2(z=10)$] 
the difference in the expansion rate and $\epsilon$ (or equivalently 
$w$) reaches several percent (see right panels in Fig.~\ref{results}), 
which is testable by the next generation of observatories such as 
Euclid~\cite{Amendola:2016saw} and ELT~\cite{Shearer:2010cn}.

In order to study the effect of the nonminimal coupling $\xi$ on the 
geometrical quantities $H(z)$ and $\epsilon(z)$, in Fig.~\ref{results2} 
we show the evolution of the Hubble parameter and $\epsilon$ as a
function of the number of e-folding 
$N \!=\! -\ln({1\!+\!z})$ for $\xi \!=\! -10^{-3}$. As in Fig.~\ref{results} 
we see that for small masses, when $m\ll H_0$ ($m\simeq 0.20H_0$) 
the difference between $\Lambda$CDM and our model is  at a 
sub-percent level and hence barely reachable by planned observatories. 
However,  for a mass comparable to the Hubble rate today 
($m\simeq 0.61H_0$), the difference becomes significant and well 
within reach of Euclid and ELT.  Comparing Figs.~\ref{results} 
and~\ref{results2} reveals that increasing $|\xi|$ seems to increase 
the difference between our model and $\Lambda$CDM. However, 
for small values of $\xi$'s ($\xi \!=\!- 10^{-3}$ in figure~\ref{results2})
 this increase is rather moderate. Note that a non-vanishing mass
 term (or a potential term) is crucial for the model to exhibit DE behaviour.
 (The case $m\!=\!0$ and $\xi\!<\!0$ is examined in~\cite{Glavan:2014uga}, and 
that model does not exhibit a late time dark energy.)

The nonminimal coupling is a relevant parameter here more from the 
point of view of relating the DE period to the primordial inflationary 
period, more precisely to the total number of e-foldings of inflation 
$N_I$. In case of zero nonminimal coupling this number has to be 
huge~\cite{Aoki:2014dqa}, $N_I\sim10^{12}$ for both cases of 
Fig.~\ref{results}, while for even rather small nonminimal couplings of 
Fig.~\ref{results2} the total number of e-foldings of inflation is much 
smaller~\cite{Glavan:2015cut}, $N_I\sim10^{5}$, which can be determined
from~(\ref{intricate relation}). Larger values (in magnitude) for nonminimal 
coupling are problematic in the sense that one has to worry about large
backreaction much earlier, during primordial inflation. Although this is an 
interesting possibility to study in the context of graceful exit from inflation, 
it is not considered in our DE model (for a recent study of the backreaction 
of a heavy nonminimally coupled scalar in de Sitter inflation 
see~\cite{Markkanen:2016aes}).
\begin{figure}[h!]
\begin{minipage}[b]{8cm}
$[m/H(z_{in})]^2 = 10^{-4}, \ \xi=0$
    \includegraphics[width=7.5cm] {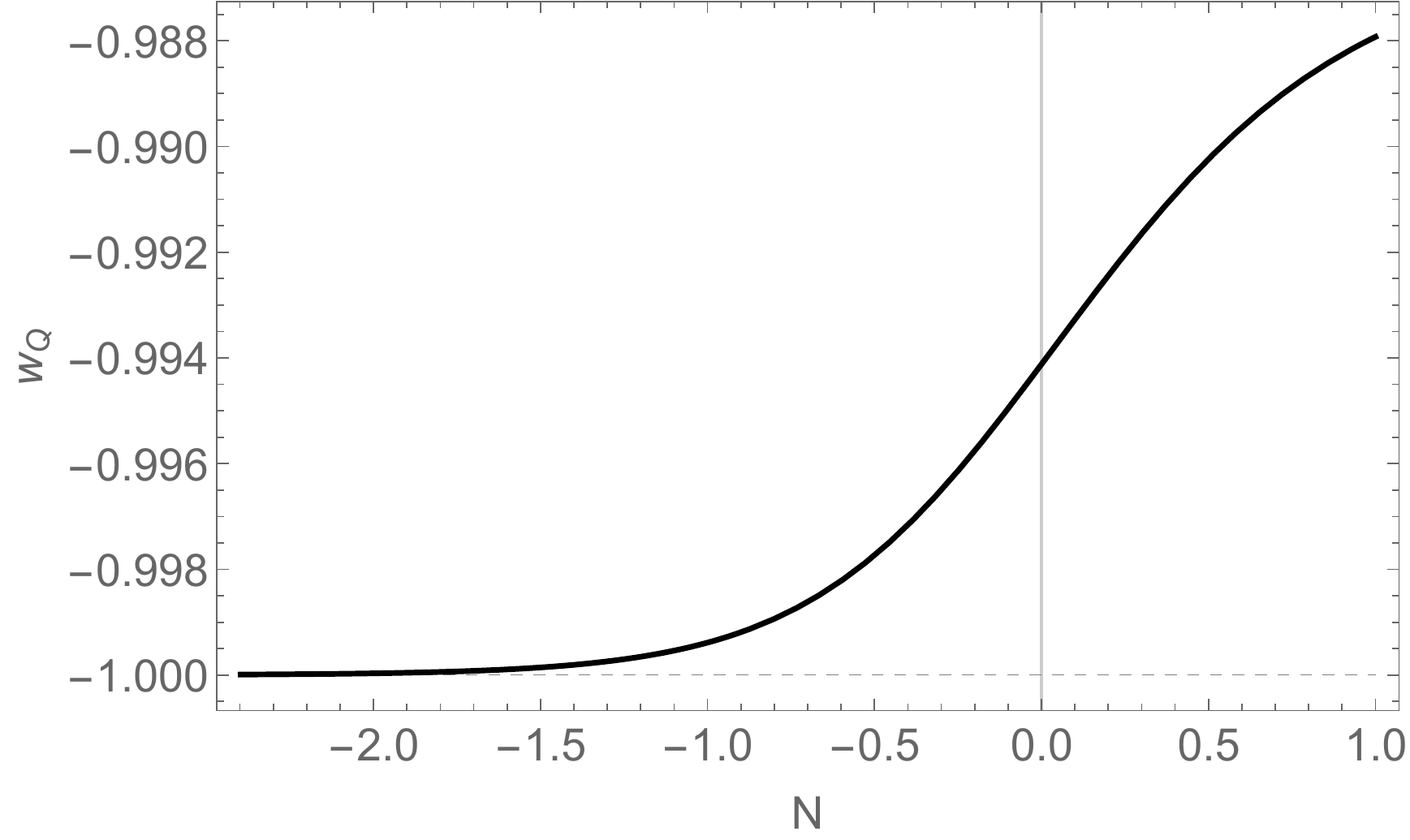}
\end{minipage}
\hfill
\begin{minipage}[b]{8cm}
$[m/H(z_{in})]^2 = 10^{-3}, \ \xi=0$
    \includegraphics[width=7.5cm] {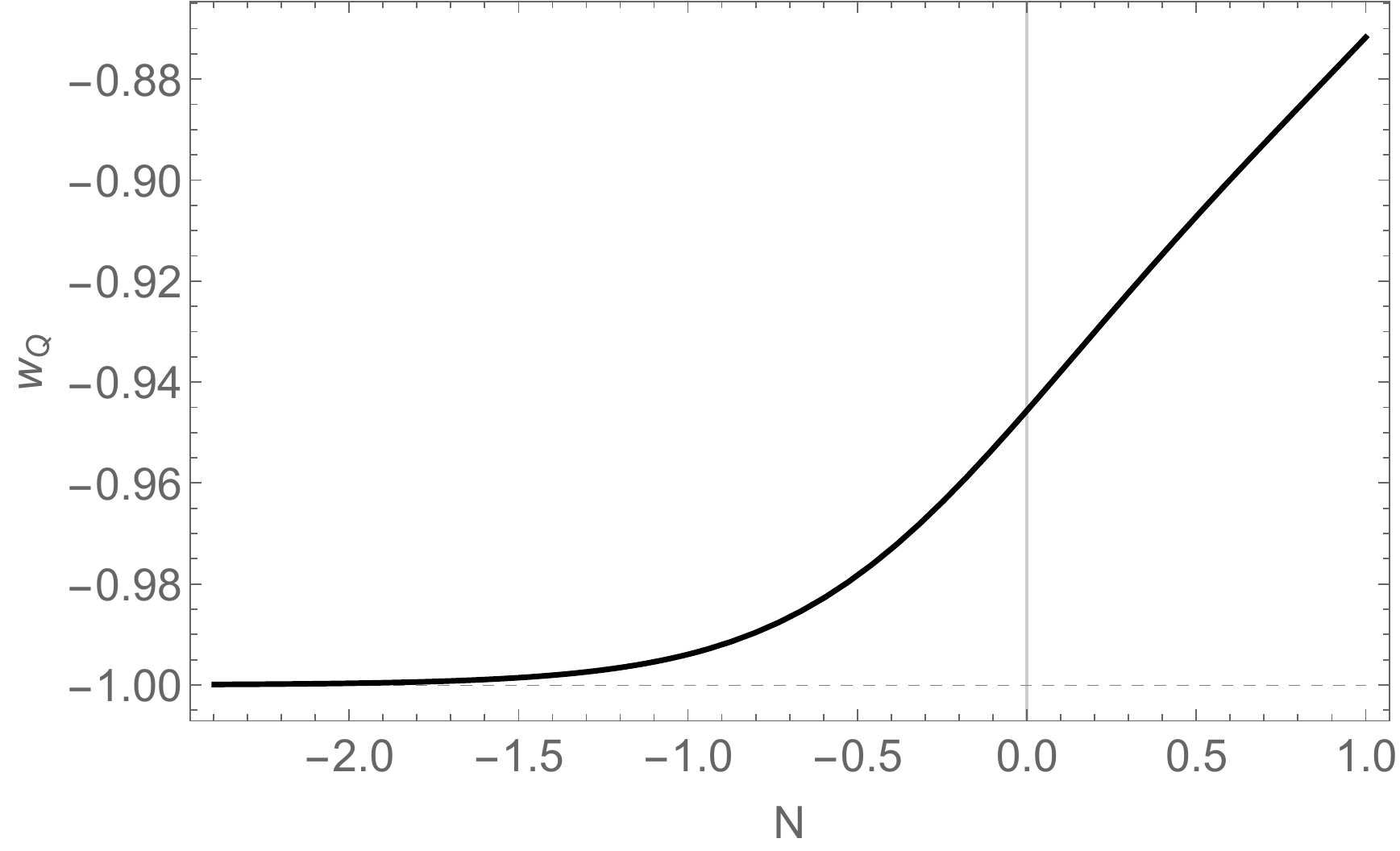}
\end{minipage}
\vskip+0.3cm
\begin{minipage}[b]{8cm}
\hspace{0.8cm} $[m/H(z_{in})]^2 = 10^{-4}, \ \xi= - 10^{-3}$
    \includegraphics[width=7.5cm] {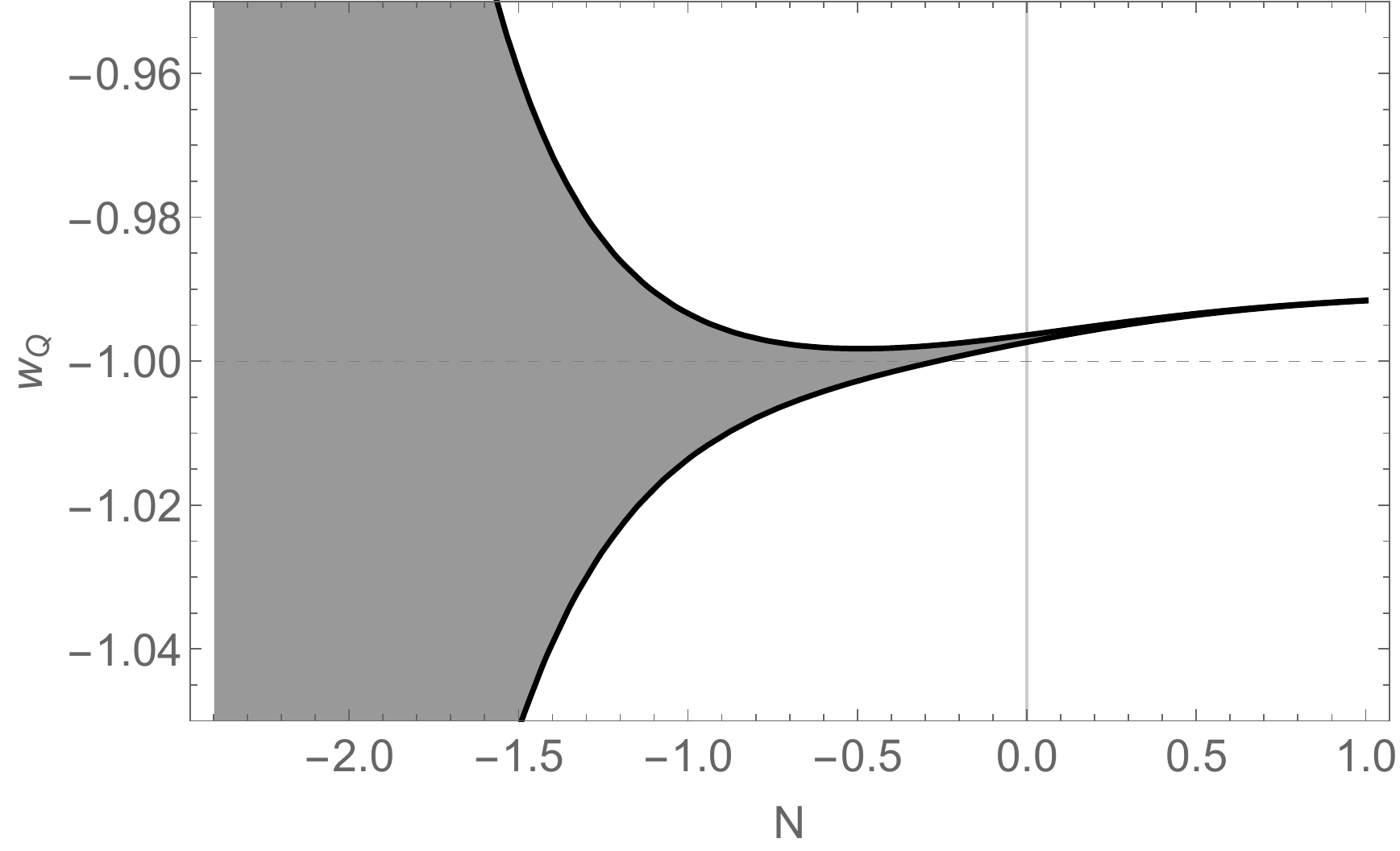}
\end{minipage}
\hfill
\begin{minipage}[b]{8cm}
\hspace{0.8cm} $[m/H(z_{in})]^2 = 10^{-3}, \ \xi= - 10^{-3}$
    \includegraphics[width=7.5cm] {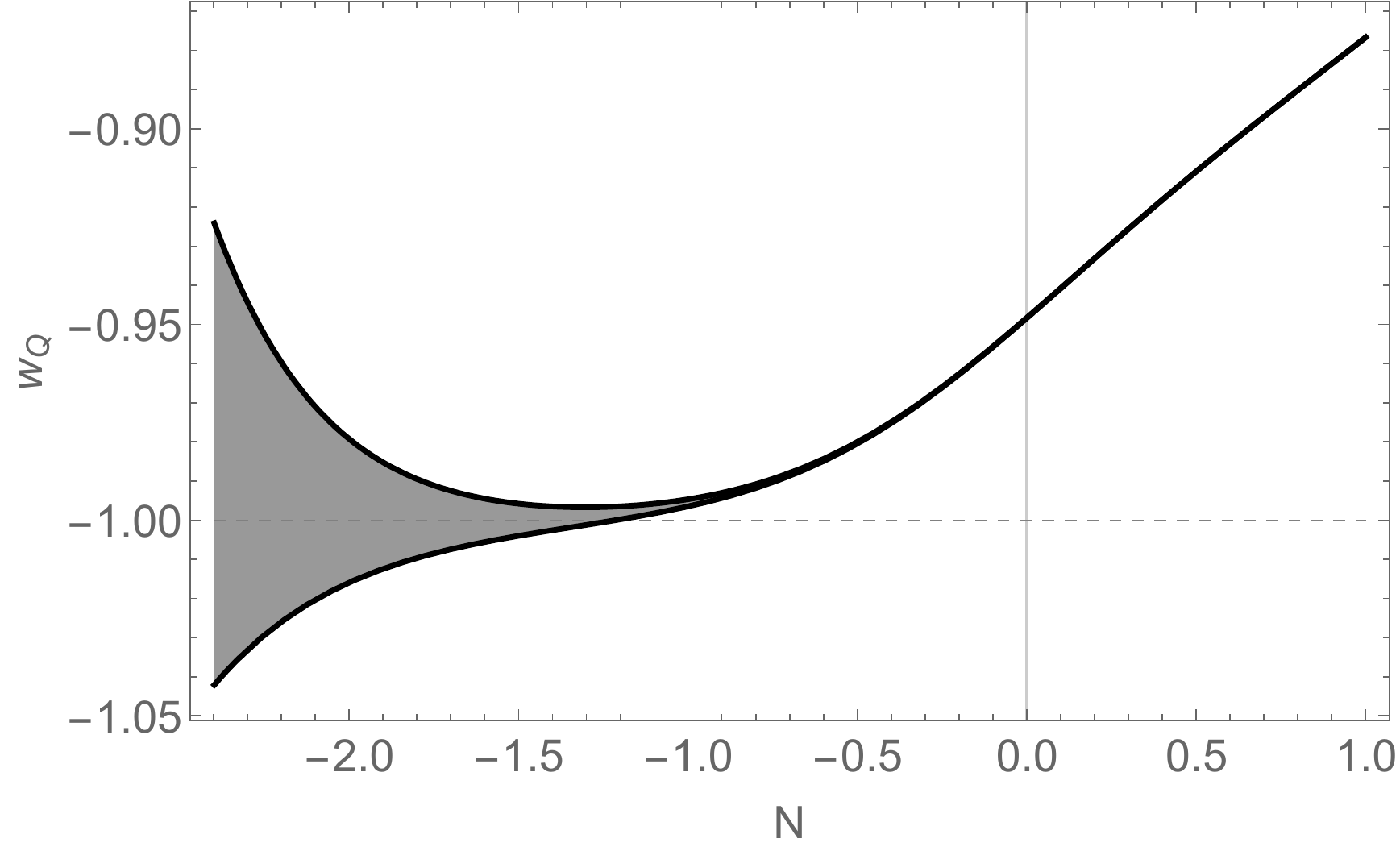}
\end{minipage}
\vskip-0.3cm
\caption{The DE equation of state parameter $w_{DE}\equiv 
w_{Q}=p_Q/\rho_Q$ for the quantum
backreaction as a function of $N\!=\!-\ln(1\!+\!z)$. To facilitate 
comparison, the parameters are chosen as in Fig.~\ref{results}
and~\ref{results2} and they are indicated above the plots.
The uncertainty in the lower two plots reflects the uncertainty related 
to the subtraction of the part of the quantum backreaction 
that scales (to a good approximation) as dark matter. Note that the uncertainty is larger 
at larger redshift. However, that does not reduces in any significant way 
testability of the model, because at large redshifts the relative contribution of dark energy is small.}
\label{results3}
\end{figure}

While Figs.~\ref{results} and~\ref{results2} nicely show the recent 
Universe dynamics in our model and compare them to those 
of $\Lambda$CDM, it is instructive to show how the dark energy 
component in our model evolves, and that is illustrated in 
Fig.~\ref{results3}. The upper two plots show that for the vanishing
nonminimal coupling ($\xi\!=\!0$), the backreaction behaves like
a cosmological constant to a very good approximation in the 
late Universe, causing the transition from the matter-dominated
to the quasi-de Sitter phase. This behaviour tends to persist until the 
Hubble rate drops 
below the mass scale of the scalar field ($m\!\sim\!H$). 
After this point the Universe transitions to a decelerated 
phase where $\epsilon$ oscillates between 0 and 3.

The lower two plots in~Fig.~\ref{results3} demonstrate the effect
of non-minimal coupling on the dark energy equation of state.
The shaded regions correspond to the uncertainty we have in estimating its
equation of state. These are introduced for the following reason -- 
due to nonminimal coupling the backreaction contains a contribution 
(whose energy density and pressure we denote as $\rho_Q^m$ and $p_Q^m$)
that to a good approximation scales like nonrelativistic matter at higher redshifts
(while the backreaction is still small), as is evident from Eq.~(\ref{mat rho p}).
It is more appropriate not to consider this contribution a part of dark energy,
but rather to group it together with the background matter fluid. Hence,
we have subtracted the contribution from the backreaction that scales away exactly
like mater, and determined the equation of state for the remaining contribution
in the backreaction fluid. The uncertainty in $w_Q$ is due to the fact that
the amplitude of this matter like contribution is computed approximately
to order $\mathcal{O}(\xi)$, and that it does not scale exactly like matter, but its
behaviour has corrections of the order of $\mathcal{O}(|\xi|N_M)$, where
$N_M$ is the number of e-foldings of the matter-dominated period.
This means we do not know exactly how much of this matter-like fluid to subtract,
which introduces sizeable uncertainties in the equation of state only for higher
redshifts, but for smaller redshifts ($z\lesssim 3$) we get -- up to sub-percent
uncertainties -- unique predictions. This implies that our model of dark energy is 
in principle testable as it gives a definite prediction for the equation of state 
parameter $w_{\rm Q}$ as a function of the redshift.


Even though we know that the subtracted matter-like contribution redshifts
like ordinary matter on cosmological scales, we do not know how it clusters.
If its clustering turns out to be insignificant, its principal effect 
would be a finite renormalization (reduction) of 
the Newton constant. In the opposite case the effect would fall under
the class of some interacting dark matter/dark energy models, the precise
nature of which requires further investigation, which is left for future
work.


As of yet dark matter has not been directly seen. However, in the near future dark matter 
may be seen in the laboratories and if that happens, we may have enough 
to infer its energy density. Since our model predicts a negative 
component that redshifts as dark matter (see Eq.~(\ref{mat rho p})), 
if it also clusters like dark matter, 
there should be more dark matter 
than what indirect observations (such as lensing) indicate. More concretely,
for $[m/H(z_{in})]^2=10^{-4}$ and $\xi=10^{-3}$ the ratio is:
$\rho_Q^m/\rho_c=-0.0687$ and 
for $[m/H(z_{in})]^2=10^{-3}$ $\xi=10^{-3}$ the ratio is
$\rho_Q^m/\rho_c=-0.0085$, which is large enough to be potentially observable
(both evaluated at redshift $z\simeq 10$). 
As far as we know, our model is the only one in the literature that exhibits this feature,
which may become a smoking gun for testing our model.

\begin{figure}[h!]
\vskip+0.7cm
\begin{minipage}[b]{8cm}
$[m/H(z_{in})]^2 = 10^{-4}, \ \xi=0$
    \includegraphics[width=7.5cm] {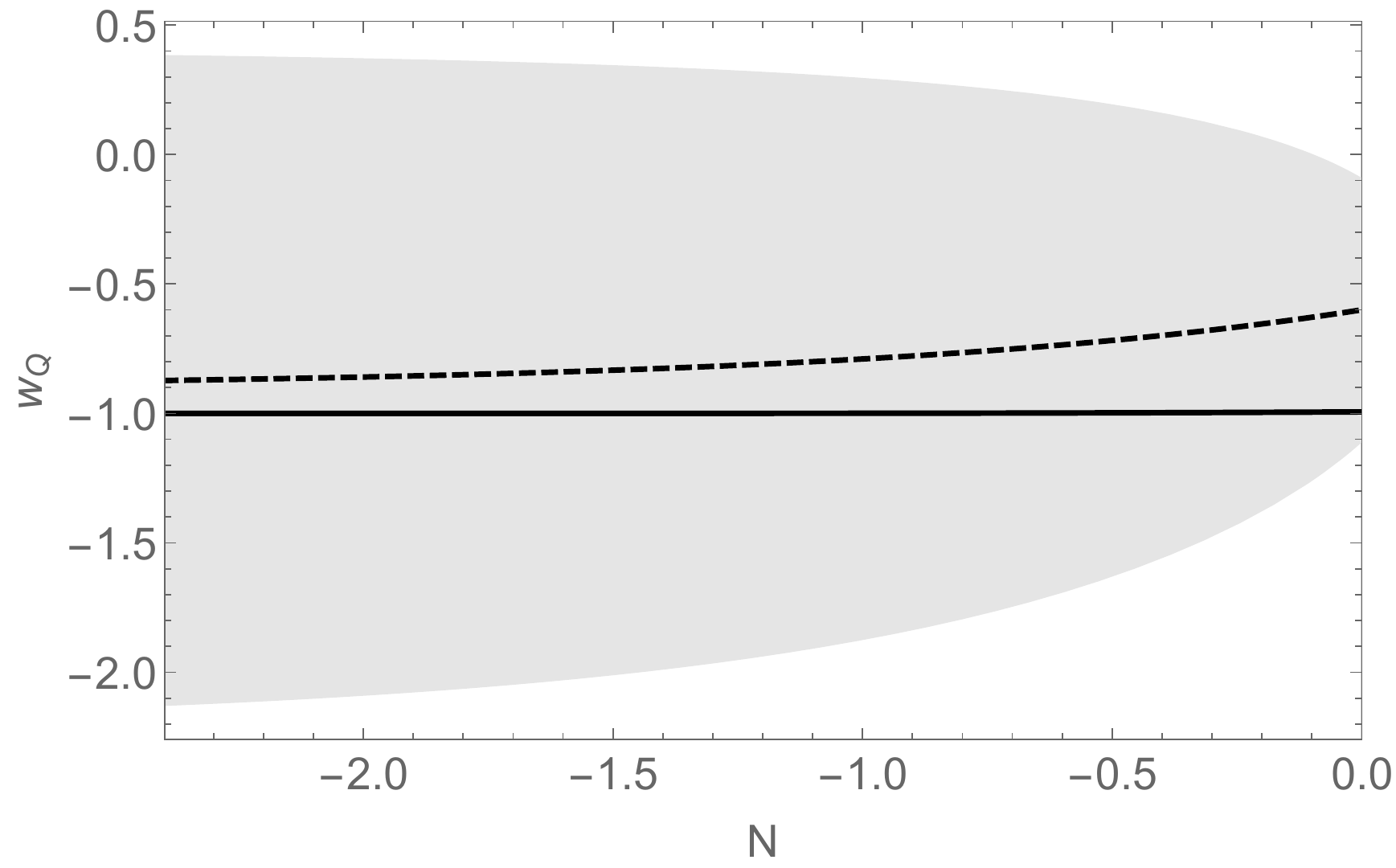}
\end{minipage}
\hfill
\begin{minipage}[b]{8cm}
$[m/H(z_{in})]^2 = 10^{-3}, \ \xi=0$
    \includegraphics[width=7.5cm] {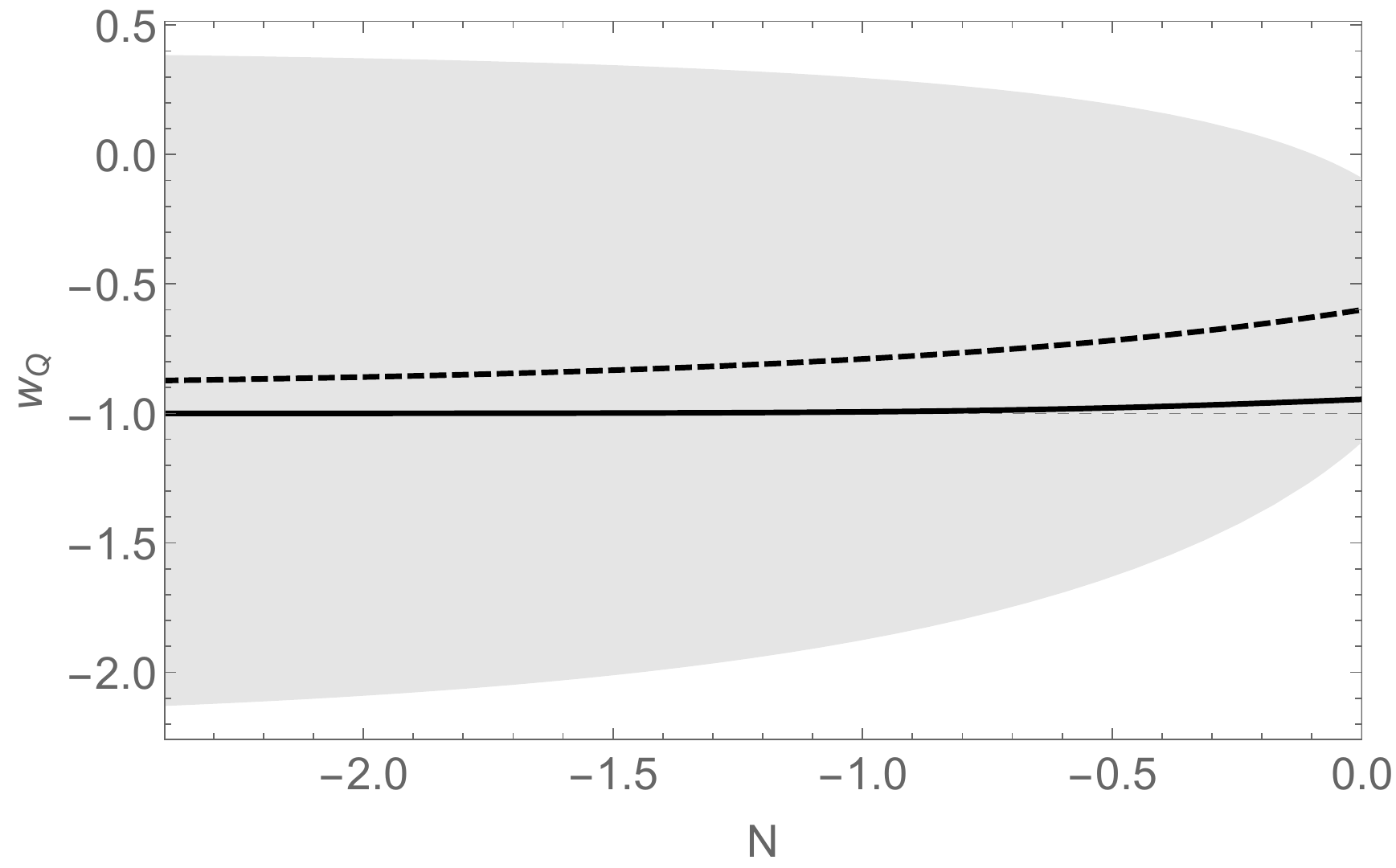}
\end{minipage}
\vskip+0.3cm
\begin{minipage}[b]{8cm}
\hspace{0.8cm} $[m/H(z_{in})]^2 = 10^{-4}, \ \xi= - 10^{-3}$
    \includegraphics[width=7.5cm] {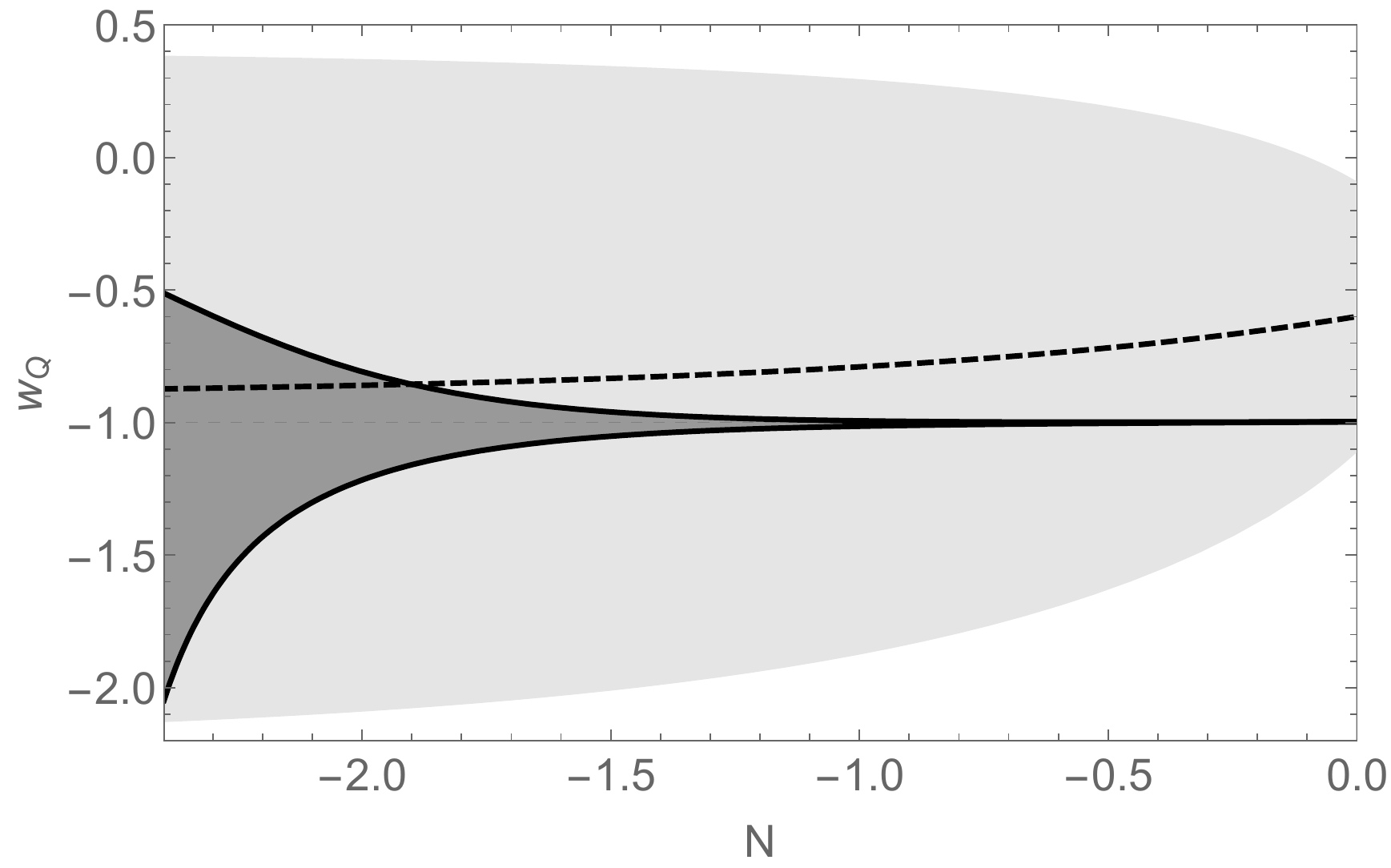}
\end{minipage}
\hfill
\begin{minipage}[b]{8cm}
\hspace{0.8cm} $[m/H(z_{in})]^2 = 10^{-3}, \ \xi= - 10^{-3}$
    \includegraphics[width=7.5cm] {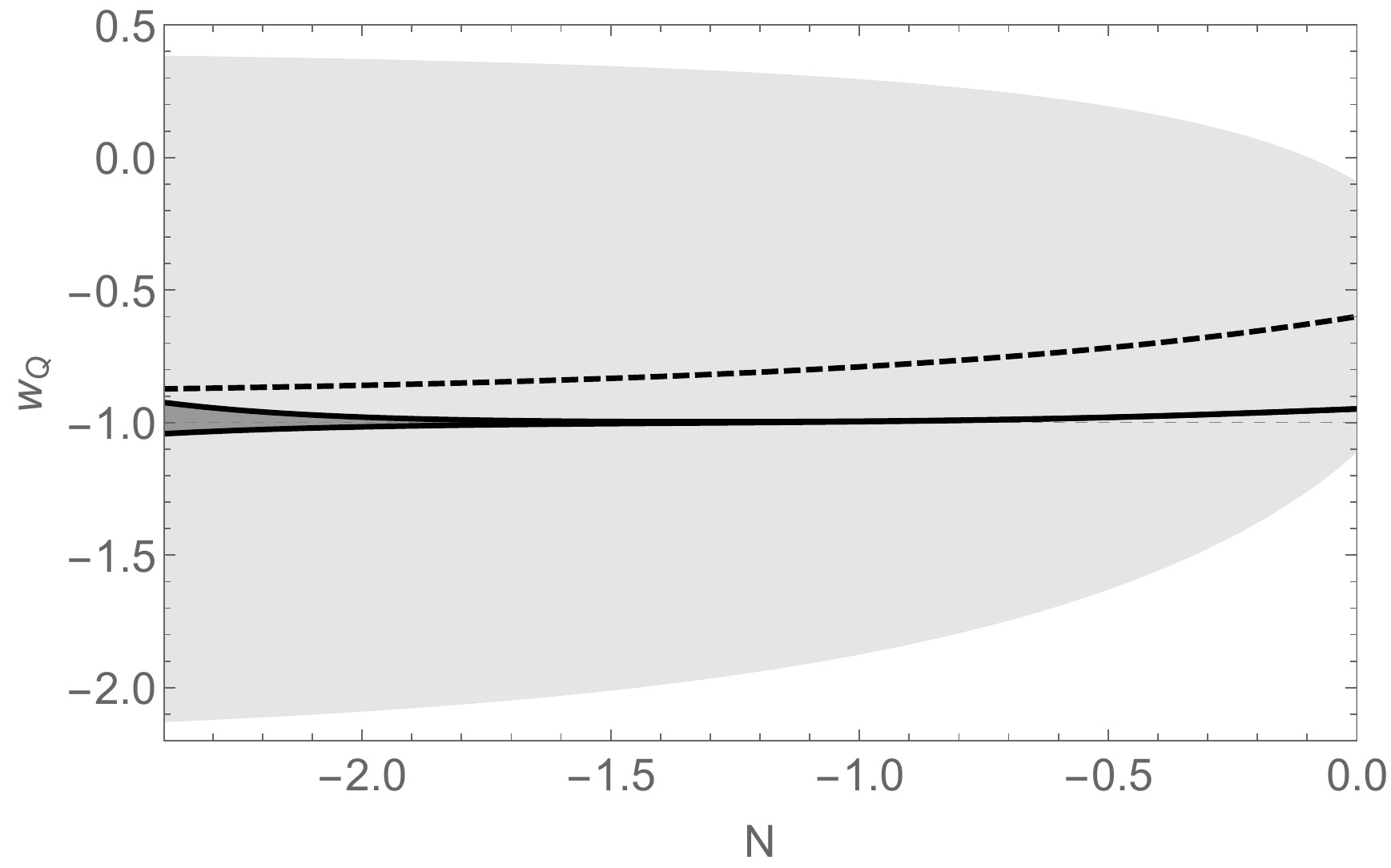}
\end{minipage}
\vskip+0.3cm
\caption{The DE equation of state parameter $w_{DE}$ as a function 
of $N \!=\! - \ln(1\!+\!z)$. 
The solid black curve shows the quantum backreaction equation of 
state parameter, $w_Q(N)$, the dashed curve shows $w_{DE}(N)$ 
in the $w_0w_a$CDM model, in which $w_{DE}(N)\!=\!w_0+(1 \!-\! e^N)w_a$,
and the shaded region is the $1\sigma$ contour corresponding to
$w_0 \!=\! -0.6\pm0.25$, $w_a \!=\! -0.3\pm0.8$. 
}
\label{mismatch}
\end{figure}
Let us now compare with current observations. The Planck collaboration
papers~\cite{Ade:2015xua,Ade:2015rim} constrain 
$w_{DE}=-1.019{+0.075\atop-0.080}$ ($68\%$~CL when the complete Planck data are 
combined with available data from lensing, baryonic acoustic oscillations (BAO) 
and $H_0$ measurements). On the other hand,  the $68\%$~CL error bars on the parameters of the $w_0w_a$CDM model [defined by, 
$w_{DE}=w_0+(1-a)w_a
=w_0+[z/(1+z)]w_a=w_0+(1-e^{N})w_a$] are much weaker,
$w_0=-0.6\pm 0.25,w_a=-0.3\pm 0.8$
 (with the Planck, BAO, SN-Ia $\&$ $H_0$ data included) and
$w_0=-0.6\pm 0.5,w_a=-1.2\pm 1.2$
 (with the Planck, BAO $\&$ redshift space distortions data included)
(see figure~4 in Ref.~\cite{Ade:2015rim}), which shows that the constraints 
$w_0$ and $w_a$ depend significantly on the data 
sets used and hence the quality of the data is still not good enough to place strong bounds on the model. 
Since these bounds are only slightly modified when the more recent 1 year DES data are 
accounted for~\cite{Abbott:2017wau}, they suffice for our purpose.
It is clear that these bounds are not precise enough to place meaningful
constraints on the dark energy models shown in Fig.~\ref{results3}.
This can be seen even clearer in Fig.~\ref{mismatch}, in which 
we observe that the contours of our dark energy model lie comfortably within one standard deviation contours 
of the $w_0w_a$CDM model (which meaningfully constrains dark energy only
at relatively small redshifts, $z\lesssim 2$, or $N\gtrsim -1$). 

 However, with the incipience of upcoming high precision observatories such as Euclid and LSST,
the constraints may become strong enough to curb our model, especially when
combined with inhomogeneous probes.


\section{Discussion and outlook}

This paper is one in a series~\cite{Glavan:2013mra,Glavan:2014uga,
Glavan:2015cut}, whose goal is to investigate viability of DE models 
based on quantum fluctuations during inflation. The attractiveness of 
these models stems from the fact that -- when compared to more 
conventional DE models -- they are amenable to additional tests. For 
example, since the origin of these models can be traced back to inflation, 
there is an intricate relationship between DE and infationary observables. 
Furthermore, since the origin of DE in these models is in quantum 
fluctuations of matter fields, in general we expect these models to 
generate inhomogeneous dark energy with calculable amount of 
inhomogeneities, which can also be used to test this class of models.

In this paper we reanalyze the dark energy model proposed in 
Ref.~\cite{Glavan:2015cut}. The model utilizes an ultra-light, 
non-minimally coupled scalar field and its dynamics is governed by the 
action~(\ref{the action}). Even though this field is a spectator field 
during inflation, quantum fluctuations naturally grow large such that, 
by the end of inflation, they typically reach super-Planckian values.
In fact, due to the assumed negative non-minimal coupling, it can take 
as little as a thousand of e-foldings to achieve this feat. This mechanism 
addresses the question of naturalness of initial conditions -- often raised 
in quintessence DE models -- by dynamically generating 
super-Planckian fluctuations from initially subhorizon quantum 
fluctuations.

In order to be able to calculate beyond the perturbative regime of 
Ref.~\cite{Glavan:2015cut}, in section III we develop a stochastic 
framework which permits us to solve the problem self-consistently, 
also in the regime where the quantum backreaction dominates the 
background dynamics. To establish the accuracy of our stochastic 
model, in section~\ref{Comparison with field theoretic calculations} 
we use the stochastic framework to calculate the one loop contribution 
to the energy density and pressure of quantum fluctuations and 
demonstrate agreement with the perturbative calculations of 
Ref.~\cite{Glavan:2015cut} in all of the relevant epochs: inflation, 
radiation and matter era. This is sufficient to establish perturbative 
accuracy of the model. The main advantage of the stochastic formulation 
is in that a self-consistent solution of semiclassical gravity comes within 
reach of even modest programmers, as the original set of 
integro-differential equations of semiclassical gravity is reduced to a 
much simpler set of ordinary differential equations with rather facile 
stochastic sources. 

A notable result of our work is that stochastic sources -- which are of 
essential importance for capturing the correct dynamics during 
inflation -- can be neglected during the post-inflationary radiation 
and matter epochs, such that the field dynamics becomes essentially 
classical. The memory of stochastic sources is still kept in the initial 
conditions for radiation era and for matter era. 

Armed with this formalism we then study how quantum inflationary 
fluctuations in our model~(\ref{the action}) evolve in time and how they 
affect the background evolution. In particular in Fig.~\ref{results}
and~\ref{results2} we compare the evolution of the Hubble rate $H(t)$ 
and its dimensionless rate of change, $\epsilon \!=\! -\dot H/H^2$ 
(which is equal to the principal slow roll parameter during inflation)
in our model with the same in $\Lambda$CDM. Our results show that, 
as the ultra-light mass $m$ is taken to be closer to, but still remains 
smaller than, the Hubble parameter today, the differences become 
significant and observable by the near future observatories such as 
Euclid and ELT. From Figs.~\ref{results3} and~\ref{mismatch} we see 
that the current data are not constraining enough to meaninfully test our 
the backreaction DE models. However, the near future data (from Euclid, SST, ELT 
and other planned observatories) will penetrate deeper into the redshift 
space and will be collecting much more data and thus will be able to constrain the 
quantum backreaction DE models.  
Next, we point out that the dependence of $w_Q=w_Q(z)$ in our quantum 
backreaction model is quite different from that in typical quintessence models~\cite{Tsujikawa:2013fta}. This owes to the rather unusual feature 
of the perturbative initial condition in matter 
era~(\ref{rho Q: matter era}--\ref{p Q: matter era}), which is in turn dictated 
by the standard (IR-regulated) Chernikov-Tagirov-Bunch-Davies initial condition 
in inflation~(\ref{mode function}). Namely, there is a significant negative 
energy component that initially scales as the dominant background 
(dark matter) component 
({\it cf.} Eqs.~(\ref{rho Q: matter era}--\ref{p Q: matter era})).
This feature is akin to negative energy  scenarios studied 
in~\cite{Prokopec:2011ce} and -- once the properties of (particle-like) dark matter 
are mapped out -- can be used to test this class of DE models.

While in this paper we only study the background evolution, keeping in 
mind the quantum inflationary origin of scalar field dark energy in our 
model, it would be of utmost importance to investigate the growth 
of structure in the visible and dark energy sector in our DE model, 
both at Gaussian and non-Gaussian levels. Namely, due to the non-minimal 
coupling the model is non-Gaussian and hence we expect it to leave distinct 
Gaussian and non-Gaussian footprints at late times on Universe's large scale 
structure. Determining these will require a suitable adaptation of the stochastic 
formalism for off-coincident correlation functions in 
inflation~\cite{Starobinsky:1994bd} or of further developments in the 
stochastic $\delta N$ formalism used in~\cite{Fujita:2014tja,Pattison:2017mbe}.

From the theoretical point of view, the model we presented here addresses 
the issue of naturalness of initial conditions, but there still remains the issue 
of fine-tuning in the form of requiring the scalar to have a tiny mass $m\le H_0$. 
This issue is not specific to our model, but is a generic feature of all classical
quintessence models.  It would be of great interest to investigate whether 
a similar effect can be reproduced within different (interacting) models, {\it via} 
a mechanism that utilises  spontaneous symmetry breaking~\cite{Frieman:1995pm} 
or 't Hooft's naturalness hypothesis~\cite{tHooft:1979rat}.


\begin{acknowledgments}
D.~G. was supported by the grant 2014/14/E/ST9/00152 of the 
Polish National Science Centre (NCN). T.~P. the acknowledges D-ITP consortium, 
a program of the NWO that is funded by the Dutch Ministry of 
Education, Culture and Science (OCW). A.~S. was supported by the 
RSF grant 16-12-10401.
\end{acknowledgments}


\end{document}